\documentclass[sigconf]{acmart}

\usepackage{booktabs} 

\usepackage{graphicx}

\usepackage{amsmath,amssymb,amsfonts}
\usepackage[T1]{fontenc}

\usepackage{numprint}

\usepackage{color, colortbl}
\definecolor{Gray}{gray}{0.9}

\usepackage{multirow}

\usepackage{acronym}
\newacro{iot}[IoT]{Internet of Things}
\newacro{iiot}[IIoT]{Industrial Internet of Things}
\newacro{manet}[MANET]{Mobile Ad Hoc Network}
\newacro{pin}[PIN]{Personal Identification Number}
\newacro{otp}[OTP]{One Time Password}
\newacro{ip}[IP]{Internet Protocol}
\newacro{mac}[MAC]{Mandatory Access Control}
\newacro{dac}[DAC]{Discretionary Access Control}
\newacro{rbac}[RBAC]{Role-Based Access Control}	
\newacro{radius}[RADIUS]{Remote Authentication Dial-In User Service}
\newacro{rfid}[RFID]{Radio Frequency IDentification}
\newacro{scada}[SCADA]{Supervisory Control And Data Acquisition}
\newacro{cots}[COTS]{Commercial Off The Shelve}
\newacro{osi}[OSI]{Open System Interconnection}
\newacro{tcp}[TCP]{Transmission Control Protocol}
\newacro{udp}[UDP]{User Datagram Protocol}
\newacro{ip}[IP]{Internet Protocol}
\newacro{plc}[PLC]{Programmable Logic Controller}
\newacro{ai}[AI]{Artificial Intelligence}
\newacro{ids}[IDS]{Intrusion Detection System}
\newacro{it}[IT]{Information Technology}
\newacro{ua}[UA]{Unified Architecture}
\newacro{cps}[CPS]{Cyber Physical System}
\newacro{rtu}[RTU]{Remote Terminal Unit}
\newacro{mtu}[MTU]{Master Terminal Unit}
\newacro{svm}[\textit{SVM}]{\textit{Support Vector Machines}}
\newacro{roc}[\textit{ROC}]{\textit{Receiver Operating Characteristic}}
\newacro{tpr}[\textit{TPR}]{\textit{True Positive Rate}}
\newacro{fpr}[\textit{FPR}]{\textit{False Positive Rate}}

\setcopyright{rightsretained}

\copyrightyear{2018}
\acmYear{2018}
\setcopyright{acmlicensed}
\acmConference[ARES 2018]{International Conference on Availability, Reliability 
and Security}{August 27--30, 2018}{Hamburg, Germany}
\acmBooktitle{ARES 2018: International Conference on Availability, Reliability 
and Security, August 27--30, 2018, Hamburg, Germany}
\acmPrice{15.00}
\acmDOI{10.1145/3230833.3232818}
\acmISBN{978-1-4503-6448-5/18/08}

\begin{document}
\title[Machine Learning for Anomaly Detection on \textit{Modbus/TCP Data}]{Evaluation of Machine Learning-based Anomaly Detection Algorithms on an Industrial \textit{Modbus/TCP} Data Set}

\author{Simon Duque Anton, Suneetha Kanoor, Daniel Fraunholz, and Hans Dieter Schotten}
\affiliation{%
  \institution{Intelligent Networks Research Group\\German Research Center for Artificial Intelligence}
  \city{Kaiserslautern}
  \country{Germany}
  \postcode{67663}
}
\email{{simon.duque_anton, suneetha.kanoor, daniel.fraunholz, hans_dieter.schotten}@dfki.de}

\renewcommand{\shortauthors}{S. Duque Anton et al.}



\begin{abstract}
In the context of the \textit{Industrial Internet of Things},
communication technology,
originally used in home and office environments,
is introduced into industrial applications.
Commercial off-the-shelf products,
as well as unified and well-established communication protocols make this technology easy to integrate and use.
Furthermore,
productivity is increased in comparison to classic industrial control by making systems easier to manage,
set up and configure.
Unfortunately,
most attack surfaces of home and office environments are introduced into industrial applications as well,
which usually have very few security mechanisms in place.
Over the last years,
several technologies tackling that issue have been researched.
In this work,
machine learning-based anomaly detection algorithms are employed to find malicious traffic in a synthetically generated data set of \textit{Modbus/TCP} communication of a fictitious industrial scenario.
The applied algorithms are \textit{Support Vector Machine (SVM)},
\textit{Random Forest},
\textit{k-nearest neighbour} and \textit{k-means clustering}.
Due to the synthetic data set,
supervised learning is possible.
\textit{Support Vector Machine} and \textit{k-nearest neighbour} perform well with different data sets,
while \textit{k-nearest neighbour} and \textit{k-means clustering} do not perform satisfactorily.\\ \par
This is a preprint of a work published in the Proceedings of the 13th International Conference on Availability, Reliability and Security (ARES 2018).
Please cite as follows: \par
S. D. Duque Anton, S. Kanoor, D. Fraunholz, and H. D. Schotten: ``Evaluation of Machine Learning-based Anomaly Detection Algorithms on an Industrial Modbus/TCP Data Set.''
In: Proceedings of the 13th International Conference on Availability, Reliability and Security (ARES 2018), ACM, 2018, pp. 41:1--41:9.
\end{abstract}

%
%



\maketitle

\section{Introduction}
\label{sec:intro}

\begin{figure}[!h]
\centering
\includegraphics[width=0.36\textwidth]{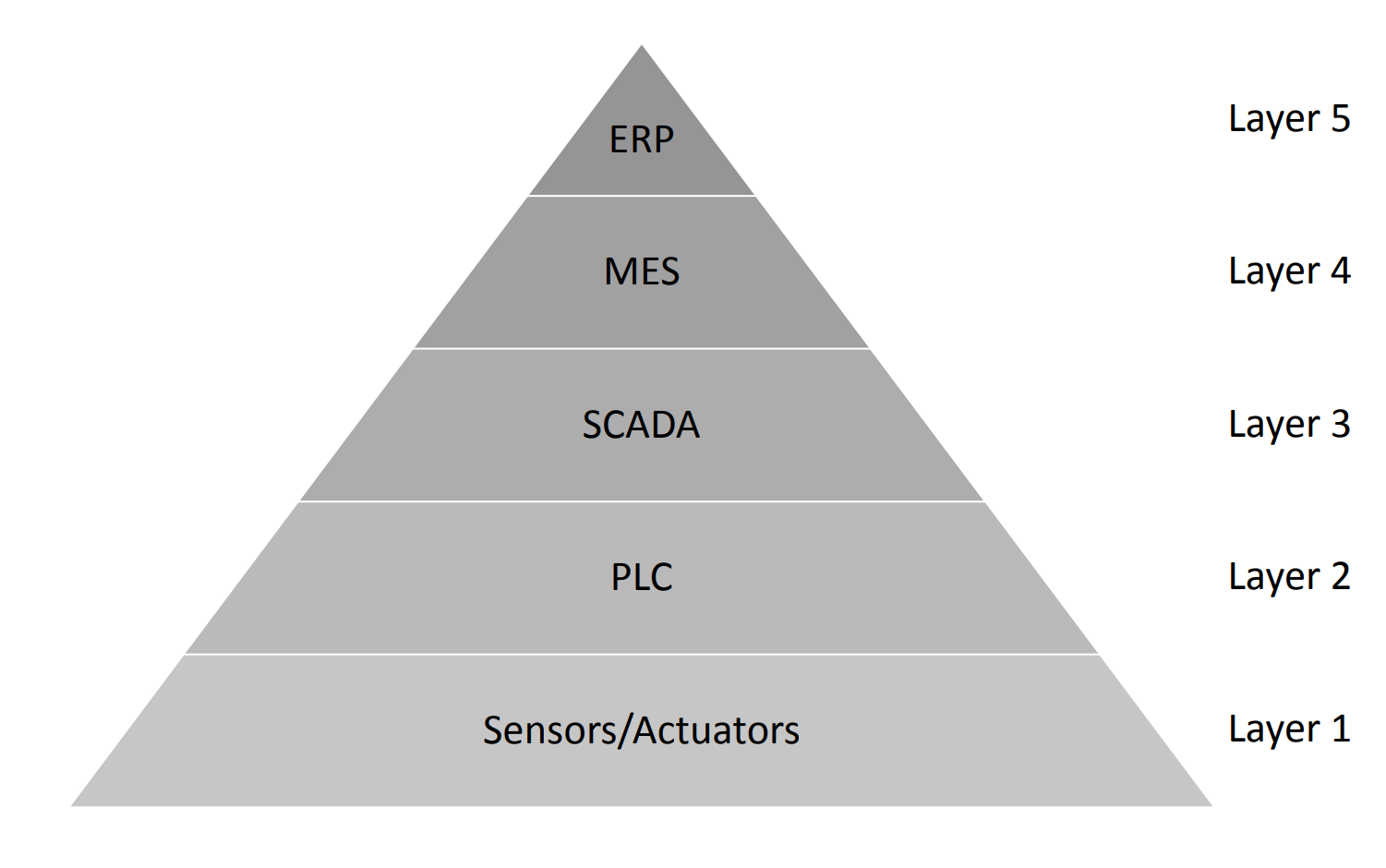}
\caption{Automation Pyramid}
\label{fig:automation_pyramid}
\end{figure}

Since the appearance of industrial control in the 1970's,
industry has been looking for ways to improve production.
At first,
hardwired sensors and actuators were employed,
followed by so-called \ac{scada} systems in the 1980's and `90's.
With the emerging computational and communication technology,
the automation pyramid,
as depicted in figure~\ref{fig:automation_pyramid},
arose.
It categorised the industrial networks according to their function:
Resource and production planning was done on the two topmost layers,
four and five.
\ac{scada} systems are located at the third layer.
\ac{plc}s can be found on the second layer,
while sensors and actuators are placed on the first layer.
This has been possible mainly due to \ac{cots} products that are interchangeable and configurable.
Many of the newly introduced network protocols either base on the Ethernet-protocol on layer 2 of the \ac{osi}-model,
such as \textit{EtherCAT}~\cite{ethercat.2018} and \textit{Modbus}~\cite{MODICONInc..1996,  ModbusIDA.2006},
or even on the \ac{tcp}/\ac{ip}-stack on layers 3 and 4 of said \ac{osi}-model.
An abundance of proprietary and open-source communication protocols,
tailored to the needs of industrial applications,
was developed.
Prominent,
\ac{tcp}/\ac{ip}-based examples are \textit{Modbus/TCP}~\cite{ModbusIDA.2006, Modbus.2012},
\textit{ProfiNET}~\cite{PROFIBUS.2017} and \textit{OPC \ac{ua}}~\cite{OPCFoundation.2017}.
After integrating communication protocols,
based on the \ac{osi}-model,
and introducing reprogrammable industrial computers,
so-called \ac{plc}s, 
industrial and automation networks are being opened to insecure networks.
The arising \ac{iiot} requires interconnectivity of networks,
accessibility and availability of resources,
also outside of trust boundaries.
A common assumption of early \ac{scada}-implementations was that networks were physically separated from public networks~\cite{Igure.2006},
breaking the trust boundaries creates a multitude of novel threats and attack vectors~\cite{Igure.2006, Zhu.2011, Duque_Anton.2017a}.
Attackers have identified industrial networks as valuable targets.
Widespread botnets provide easy opportunities to probe and capture unprotected \ac{iot}-devices~\cite{Fraunholz.2017}.
This novel threat landscape necessitates new approaches for intrusion detection and attack prevention.
Machine learning technologies,
and methods of artificial intelligence have proven that they provide vast capacities for solving problems that were hard to solve otherwise.
\\ \par
The remainder of this work is structured as follows:
Related work in artificial intelligence and machine learning for intrusion detection in industrial networks is presented in section~\ref{sec:sota}.
After that,
the employed data set is presented in section~\ref{sec:dissecting}.
First,
\textit{Modbus/TCP} is presented as a protocol.
Second,
the data set is evaluated,
and third,
features are derived.
The algorithms for anomaly detection and their application on the data set is described in section~\ref{sec:anom_detec}.
Results are discussed in section~\ref{sec:res}.
Finally,
a conclusion is drawn in section~\ref{sec:conclusion}.

\section{Related Work}
\label{sec:sota}
Intrusion detection in office \ac{it} environments is a well-researched and well-established area.
Tools,
such as \textit{Bro}~\cite{Bro.} and \textit{Snort}~\cite{Snort.},
are commonly used and maintained by a widespread user base.
They allow for easy integration of custom rules and make efficient firewalls and systems for detecting attackers and intrusions.
The same holds for data sets of host and network traffic.
There are numerous data sets to train and test \ac{ids} appliances and machine learning methods,
one of the most famous being the `99 KDD Cup data set~\cite{KDD.1999}.
As this work focuses on industrial applications of anomaly detection,
\acp{ids} for office applications,
as well as data sets with home- and office-based network traffic are not considered here.
\\ \par
One of the most important aspects in the development of novel intrusion detection approaches is a sound data set to test the system and verify the findings.
As mentioned,
there is an abundance of such data sets for home and office network traffic,
while industrial network traffic is still relatively rare.
One recent data set is presented by \textit{Lemay and Fernandez}~\cite{Lemay.2016}.
They propose an architecture for a traffic simulation environment,
based on commonly used \textit{ModbusTCP} tools and sandbox environments.
They also published a data set into which malicious traffic has been introduced.
This data set was analysed in this work in order to evaluate the effectiveness of machine learning for industrial intrusion detection.
Other than that,
\textit{Wang et al.} propose a simulation environment for \ac{scada} security analysis~\cite{Wang.2010}.
Their framework allows setting up \textit{OPC \ac{ua}} components,
including sensors and actuators,
in a simulation in order to test and verify security solutions.
Furthermore,
\textit{Siaterlis et al.} propose a testbed for the effects of cyber attacks on \ac{cps}~\cite{Siaterlis.2013}.
The testbed is based on an \textit{Emulab}~\cite{Emulab.} emulation environment and is capable of monitoring the impact of an attack on a production system.
\textit{Genge et al.} follow a similar approach~\cite{Genge.2012}.
They present an adaptable testbed that is capable of emulating different industrial production scenarios.
These scenarios can then be attacked with real malware and the effects can be evaluated.
Their testbed is based on \textit{Emulab}~\cite{Emulab.} as well, 
with a real-time connection simulator \textit{Simulink}~\cite{Simulink.}.
\textit{Seidl} designend a \textit{Python}~\cite{Python.}-environment that simulates user-defined industrial behaviour called \textit{VirtuaPlant}~\cite{Seidl.2015}.
This simulation can then be introduced to attacks and malware.
\\ \par
In addition to simulation environments,
there are also data sets available in order to train intrusion and anomaly detection algorithms.
\textit{Morris and Gao} present several files containing sets of industrial control system traffic~\cite{Morris.2014, Morris.}.
As malicious traffic is introduced into these data,
algorithms can be trained to detect traffic of malware.
\\ \par
Apart from the issue of obtaining sound and plausible data,
there is an abundance of algorithms for anomaly detection that could be employed in order to detect intrusions.
Intrusion detection as a concept,
including a formal model,
was originally presented by \textit{Denning} in 1987 in the context of the growing influence of computer systems and networks~\cite{Denning.1987}.
The applications of anomaly detection mechanisms for network intrusion detection are discussed in several surveys~\cite{Garcia-Teodoro.2009, Bhuyan.2014};
\textit{Yang et al.} give a brief introduction of these techniques for the domain of \ac{scada} systems~\cite{Yang.2006}.
\textit{Meshram and Haas} published a roadmap of machine learning based anomaly detection in industrial networks,
containing a simulation environment,
as well as a semantic description of content~\cite{Meshram.2016}.
\textit{Kleinman and Wool} present a model of the \textit{Siemens S7} protocol for intrusion detection and forensics~\cite{Kleinmann.2014}.
Critical infrastructures and industrial environments are considered in the work of \textit{Hadziosmanovic et al.}~\cite{Hadziosmanovic.2011}.
A framework that detects malicious and undesired actions is presented.
Deriving features that can be used to distinguish valid from malicious traffic is the first step in applying an intrusion detection algorithm.
\textit{Mantere et al.} look into the derivation of features from \ac{ip} traffic in an industrial environment~\cite{Mantere.2013}.
Deterministic properties of industrial control systems,
as well as the usability of this feature for anomaly detection in an industrial environment,
is researched by \textit{Hadeli et al.}~\cite{Hadeli.2009}.

\section{Dissecting the Data set}
\label{sec:dissecting}
In this section,
the data set is described.
First, 
a general introduction to the \textit{Modbus} protocol is given in subsection~\ref{ssec:intro}.
After that,
the data set used in the course of this work,
presented by \textit{Lemay and Fernandez}~\cite{Lemay.2016},
is described in subsection~\ref{ssec:descr}.
Finally,
the features that have been extracted and derived are presented in subsection~\ref{ssec:features}.

\subsection{An Introduction to \textit{Modbus}}
\label{ssec:intro}
\begin{figure}[!h]
\centering
\includegraphics[width=0.375\textwidth]{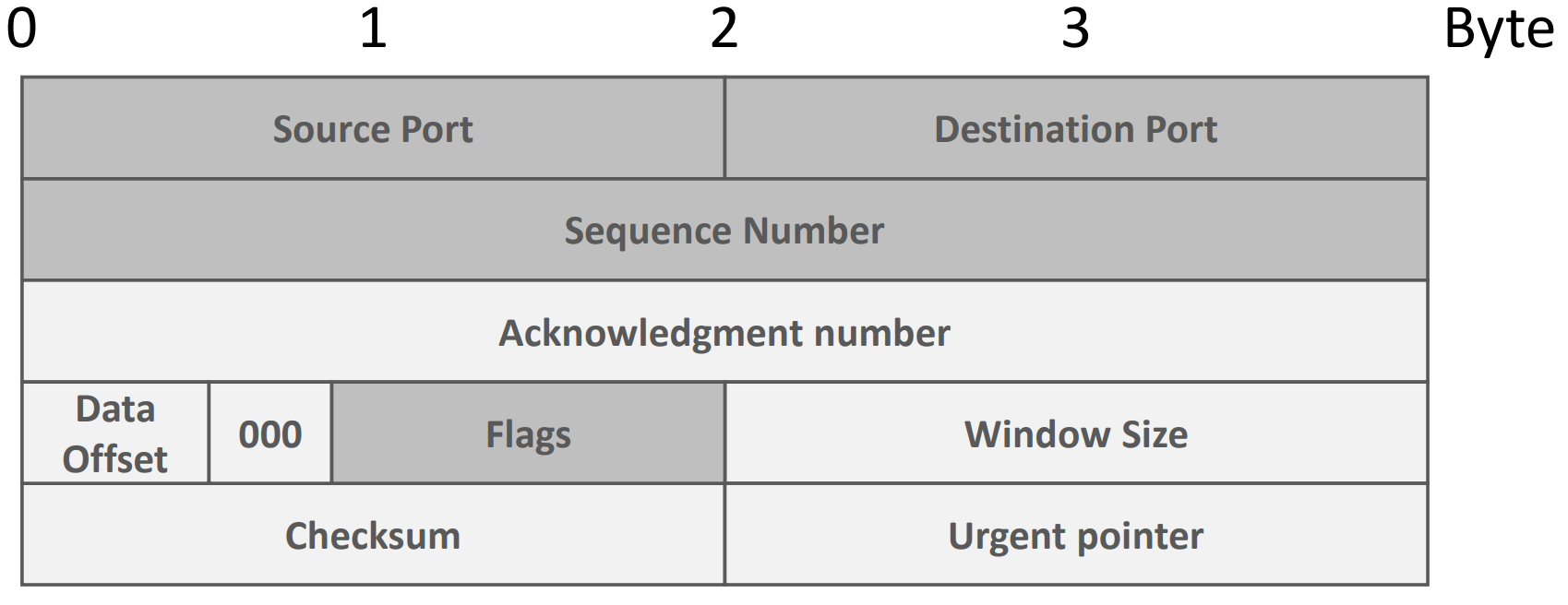}
\caption{TCP Frame Structure}
\label{fig:tcp}
\end{figure}

\begin{figure}[!h]
\centering
\includegraphics[width=0.375\textwidth]{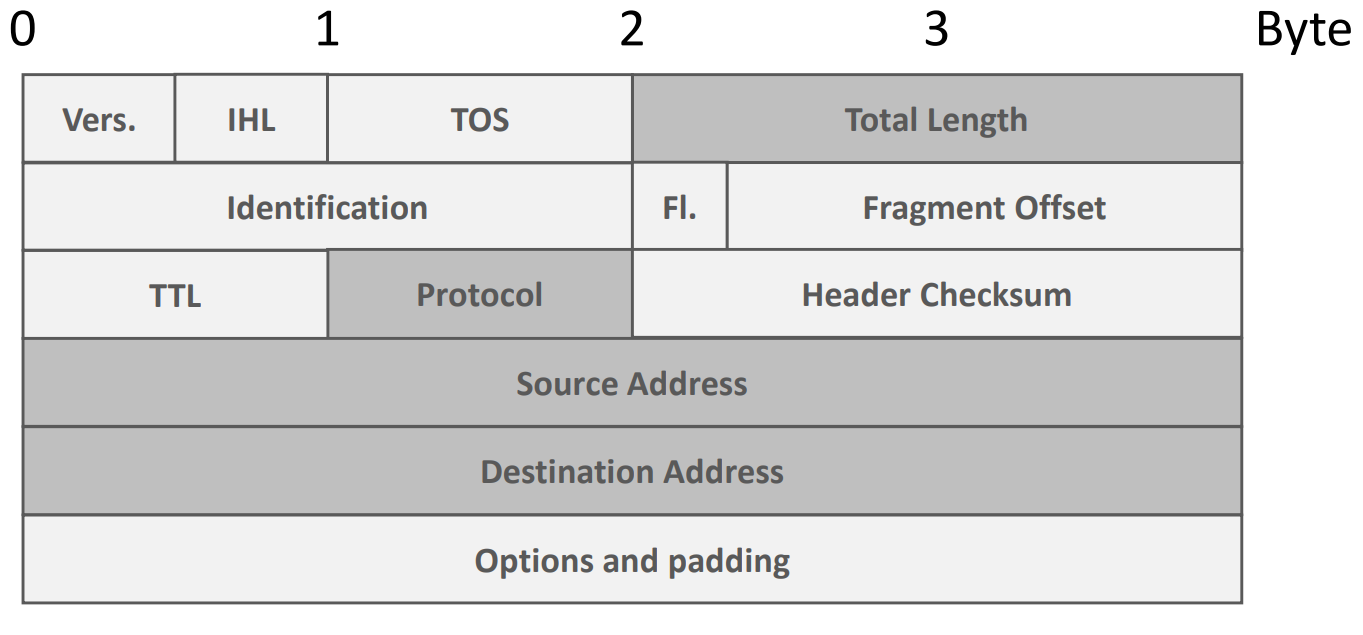}
\caption{IP Frame Structure}
\label{fig:ip}
\end{figure}

\begin{figure}[!h]
\centering
\includegraphics[width=0.475\textwidth]{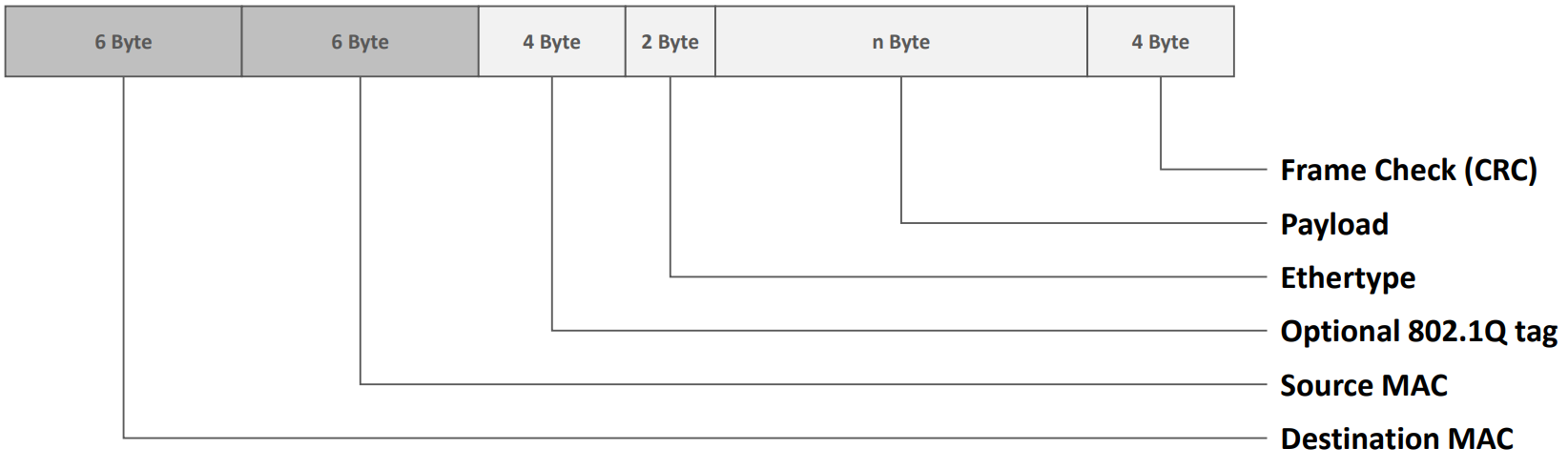}
\caption{Ethernet Frame Structure}
\label{fig:ethernet}
\end{figure}

\textit{Modbus} is a communication protocol for serial communication among \acp{plc} and \acp{rtu}.
It has been developed in 1979 by \textit{Schneider-Electric},
formerly known as \textit{Modicon}~\cite{Schneider-Electric.2017}.
It has become a de-facto standard communication protocol for industrial communication~\cite{Drury.2009}.
There are several versions of \textit{Modbus} available,
the most noteworthy are listed in table~\ref{tab:modbus_flavours}.
\begin{table}[!h]
\renewcommand{\arraystretch}{1.3}
\scriptsize
\centering
\caption{An Overview of the Most Used \textit{Modbus} Versions}
\label{tab:modbus_flavours}
\begin{tabular}{ l l }
\toprule
\textbf{Version} & \textbf{Description}  \\
\textit{Modbus RTU} & Serial communication via \textit{RS-232} connector to \\
& connect \ac{plc}s with \ac{rtu}s  \\
\textit{Modbus ASCII} & Same connector as above, but instead of binary \\
& coding, ASCII-encoded characters are used  \\
\textit{Modbus TCP/IP} & Communication based on the \ac{tcp}/\ac{ip} protocol stack  \\
& Same as above, but including a checksum in the \\
\textit{Modbus over TCP/IP} & payload, in addition to error correction mechanisms  \\
& provided by layers 1 to 4 of the \ac{osi} model  \\
\bottomrule
\end{tabular}
\end{table}
In \textit{Modbus/TCP},
communication is encapsulated in a \ac{tcp}/\ac{ip} packet,
as shown in figures~\ref{fig:tcp} and~\ref{fig:ip}.
They are transmitted via ethernet,
which follows the structure depicted in figure~\ref{fig:ethernet}.
All dark gray fields were employed as features in this work.
Most \textit{Modbus/TCP} messages contain commands regarding reading and writing coils or registers.
In analogy to analogue control automation,
one bit registers are called \textit{coils}.
Multi-bit registers are called \textit{registers}.
\textit{Modbus} slaves poll their communication and either set the data as new input for their registers or load information into a register for a master to read it.
They then respond to the request.

\subsection{Description of the Data Set}
\label{ssec:descr}
\textit{Lemay and Fernandez} simulated a controller network,
consisting of a number of \ac{mtu} and of a number of controllers.
The controllers control a simulated physical system with a \numprint{12000} Volt power source,
as well as main- and sub-branch cut-off breakers.
In this scenario,
different data collections have been performed.
An exhaustive description can be found in their work~\cite{Lemay.2016}.
Regular polling and manual operation are part of these data sets,
as these actions occur in productive systems in this form.
After collection,
malicious activities, 
generated by state of the art penetration testing tools such as \textit{metasploit}~\cite{metasploit.},
are introduced.\\ \par
This is one of the most prevalent drawbacks of the employed data set:
according to \textit{Morris and Gao} there are several different groups of \textit{Modbus}-based attacks~\cite{Morris.2014}.
Unfortunately,
none of these is introduced into this data set.
Instead,
attacks that are are also common in home and office-based penetration testing are introduced. 
Unfortunately,
this does not mimic a wide range of attacks that could be employed against industrial applications.
It does,
however,
mimic the timing behaviour and the rate of packets per time unit,
which is a good distinguishing factor for attacks.
\\ \par
In this work,
three data sets,
henceforth called \textit{DS1} to \textit{DS3},
were used for testing the algorithms:

\begin{itemize}
\item \textit{DS1: Moving\_two\_files\_Modbus\_6RTU}: Regular traffic between one \ac{mtu} and six \ac{rtu} during three minute interval, \numprint{3319} packets captured, contains \numprint{75} malicious instances
\item \textit{DS2: Send\_a\_fake\_command\_Modbus\_6RTU\_with\_operate}: Regular traffic between  one \ac{mtu} and six \ac{rtu} during 10 minute interval, \numprint{11166} packets captured, contains 10 malicious instances
\item \textit{DS3:} A combination of eight data sets, four of which do and four of which do not contain malicious activitiy, \numprint{365906} packets overall, contains 206 malicious instances
\end{itemize}

\textit{DS3} addresses a common problem in real-world intrusion detection with machine learning:
In order to train the algorithm, 
a normal condition of the system has to be derived,
deviations from which have to be recognized as anomalies.
A common practice is to monitor the behaviour of the productive system for a certain time under the assumptions that it does not contain malicious traffic.
There are two issues, 
however.
First,
in productive systems, 
you can never be sure that there is no malicious traffic.
It is just highly unlikely.
Second,
the recognition of anomalies based on normal behaviour can be difficult,
as the user usually does not know the characteristics that have most impact on the algorithm.
The effect of these limitations are evaluated in this work by mixing different kinds of traffic,
even traffic with no malicious content.
Due to the synthetic nature of this traffic,
one can be sure that it is non-malicious.

\subsection{Feature Extraction}
\label{ssec:features}
The first step in anomaly detection and data mining is the determination of relevant features.
These features can be used to describe the data instances with respect to a given goal;
the goal in the given case is to determine instances that differ significantly from the common,
productive behaviour.
Hence,
features that are suited to describe the normal behaviour of the system are needed.
In general,
there are two different kinds of features:
Basic and derived features.
Basic features are already present within the data.
In the given case,
they are contained within the protocol headers.
Network traffic,
for example,
contains source and destination addresses,
lengths,
time stamps and other features.
An exhaustive list containing the 14 basic features of this data set can be found in table~\ref{tab:basic_features}.
Two features are derived from the ethernet header as shown in figure~\ref{fig:ethernet},
four features each are obtained from \ac{tcp} and \ac{ip} header,
as shown in figures~\ref{fig:tcp} and~\ref{fig:ip},
two features from \ac{udp} headers respectively and two features from the capturing tool,
namely arrival time and information about broken packets.
\begin{table}[!h]
\renewcommand{\arraystretch}{1.3}
\centering
\scriptsize
\caption{The Basic Features Considered in this Work}
\label{tab:basic_features}
\begin{tabular}{ l l }
\toprule
\textbf{Feature} & \textbf{Description}  \\
\textit{frame.number} & Sequential number of packet \\
\textit{frame.time} & Arrival time of packet with millisecond accuracy  \\
\textit{eth.src} & Ethernet source address (MAC)  \\
\textit{eth.dst} & Ethernet destination address (MAC)  \\
\textit{ip.src} & \ac{ip} source address  \\
\textit{ip.dst} & \ac{ip} destination address  \\
\textit{ip.proto} & Transport Layer protocol  \\
\textit{frame.len} & Length of \ac{ip}-packet in bytes  \\
\textit{tcp.flags} & Control bits of \ac{tcp}-packet  \\
\textit{tcp.srcport} & Port number of source in \ac{tcp} connection  \\
\textit{tcp.dstport} & Port number of destination \ac{tcp} connection  \\
\textit{udp.srcport} & Port number of source UDP connection  \\
\textit{udp.dstport} & Port number of destination UDP connection  \\
\textit{tcp.analysis.lost\_segment} & A label set if there is a lost segment \\
\bottomrule
\end{tabular}
\end{table}
Derived features result from the combination of basic features and can often only be derived from sequences of packets,
e.g. the number of packets per time unit.
Given the time stamp and the number of bytes of each packet,
for example,
the transmitted amount of bytes per second can be calculated.
A list of nine derived features generated from this data set can be found in table~\ref{tab:derived_features}.
\begin{table}[!h]
\renewcommand{\arraystretch}{1.3}
\centering
\scriptsize
\caption{The Derived Features Considered in this Work}
\label{tab:derived_features}
\begin{tabular}{ l l }
\toprule
\textbf{Feature} & \textbf{Description}  \\
\textit{frame.time.min} & Time of frame in minutes \\
\textit{packets\_per\_minute} & Number of packets per minute  \\
\textit{frame.time.sec} & Time of frame in seconds  \\
\textit{packets\_per\_sec} & Number of packets per second  \\
\textit{packets\_per\_ip.dst} & Number of packets per unique destination-\ac{ip}  \\
\textit{stats.packets\_per\_proto} & Number of packets per protocol  \\
\textit{max\_packets} & Maximum number of packets per second  \\
\textit{min\_packets} & Minimum number of packets per second  \\
\textit{mean\_packets} & Mean number of packets per second  \\
\bottomrule
\end{tabular}
\end{table}
The impact of each feature on the prediction can be calculated.
In order to do so,
the decrease of accuracy of the prediction is evaluated.
The higher the decrease,
the more important the feature.
Another metric for the importance of a feature is the decrease in Gini index.
The Gini index describes the pureness of a data set,
split according to a given feature~\cite{Rokach.2005}.
The higher the decrease in Gini index,
the more a feature is suited to split a data set into anomalous and non-anomalous.
\\ \par
\begin{figure}[!h]
\centering
\includegraphics[width=0.375\textwidth]{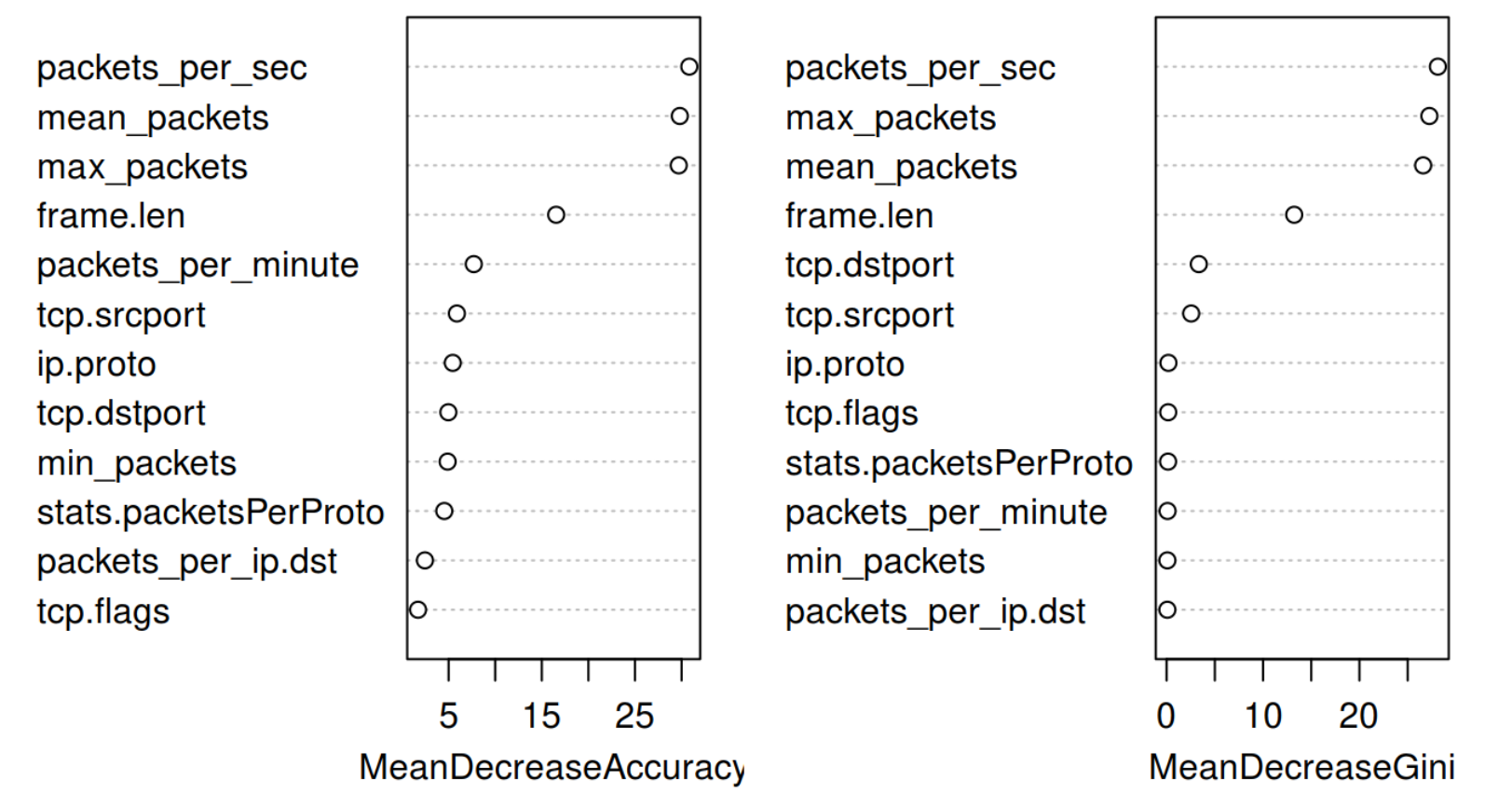}
\caption{Importance of Features for \textit{svm} in \textit{DS1}}
\label{fig:svm_ds1}
\end{figure}
\begin{figure}[!h]
\centering
\includegraphics[width=0.375\textwidth]{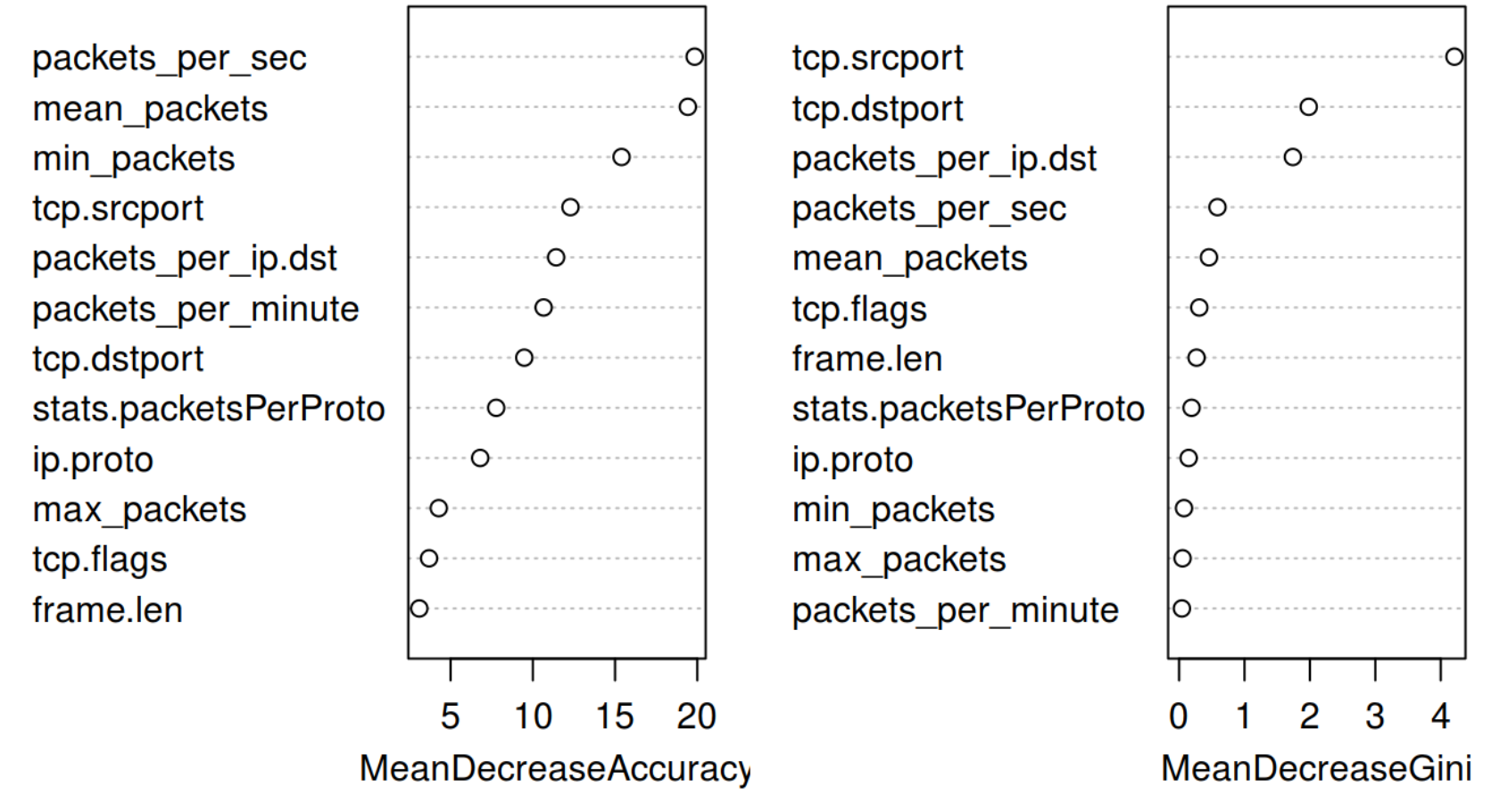}
\caption{Importance of Features for \textit{svm} in \textit{DS2}}
\label{fig:svm_ds2}
\end{figure}
\begin{figure}[!h]
\centering
\includegraphics[width=0.375\textwidth]{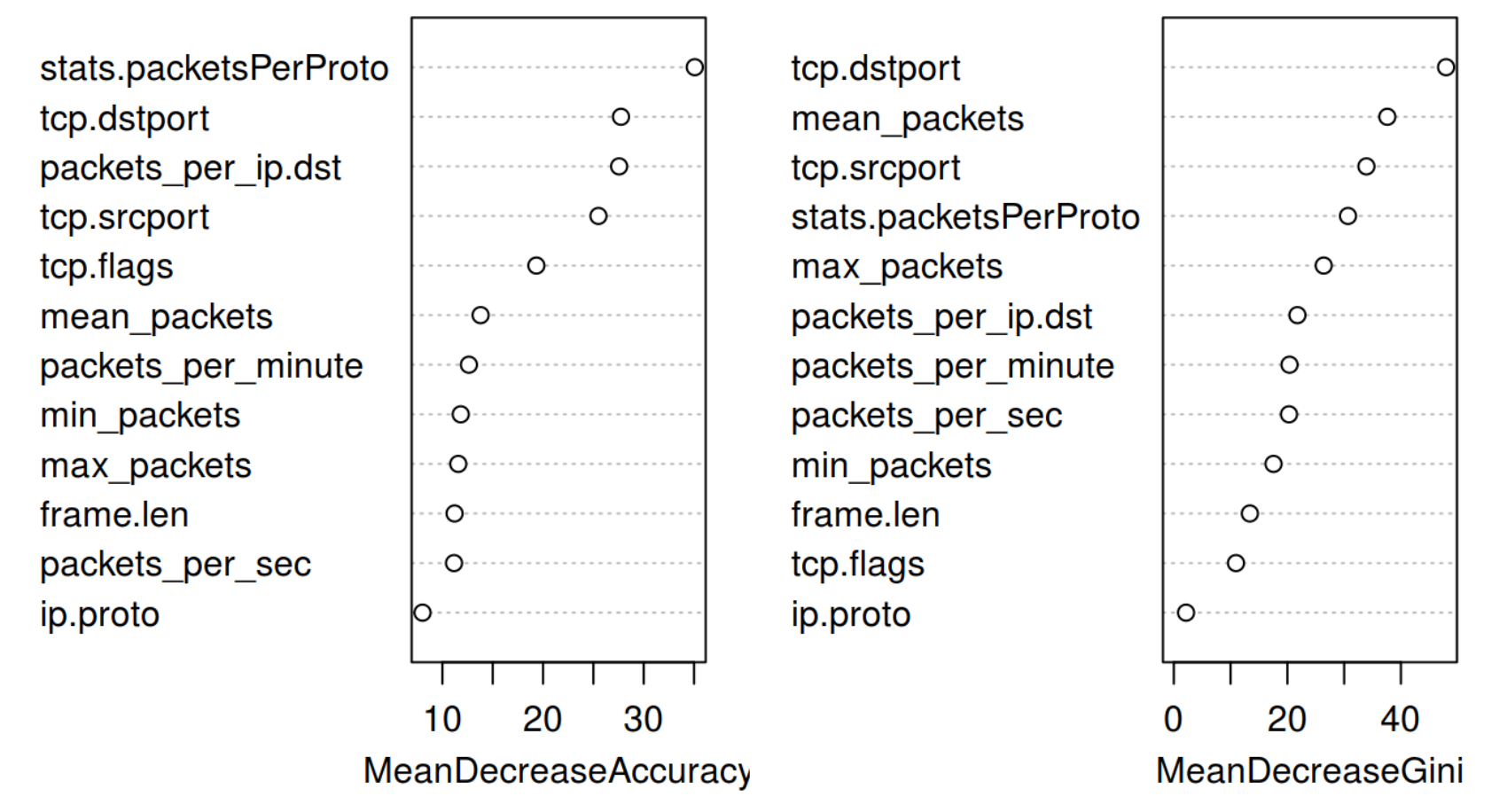}
\caption{Importance of Features for \textit{svm} in \textit{DS3}}
\label{fig:svm_ds3}
\end{figure}

The \textit{packets\_per\_second},
\textit{mean\_packets} and \textit{max\_packets} are the features with the highest impact on the result for data sets \textit{DS1} and \textit{DS2} as shown in figures~\ref{fig:svm_ds1} and~\ref{fig:svm_ds2}.
The fact that all of them are derived features underlines the importance of feature engineering.
Due to the characteristic of \textit{DS3},
consisting of different kinds of traffic,
another feature importance occurs,
as depicted in figure~\ref{fig:svm_ds3}. 
The \ac{tcp} destination port are of importance in differentiating,
as well as the number of pakets per protocol and per destination \ac{ip}.
Furthermore, 
the \ac{tcp} source port distribution and the mean number of packets per second are important.
In this scenario,
some basic features are of high importance for the anomaly detection.
A sound understanding of the scenario and application area therefore is of the essence.

\section{Anomaly Detection in \textit{Modbus} Data}
\label{sec:anom_detec}
In this section,
the application of four different machine learning algorithms,
namely 
\ac{svm}, 
\textit{Random Forrest},
\textit{k-nearest neighbour} and \textit{k-means clustering},
is described.
Those algorithms are used to find outliers in the three data sets \textit{DS1},
\textit{DS2} and \textit{DS3},
described in subsection~\ref{ssec:descr},
using the features presented in subsection~\ref{ssec:features}.
At first, 
the data sets are split into 70\% and 30\%,
as well as 80\% and 20\% respectively for cross-validation.
The split values used depend on the quality of the cross-validation,
the one providing better results is chosen.
In this work,
an  80\%/20\% has only been chosen for the \textit{k-nearest neighbour} as described in subsection~\ref{ssec:knn}.
The larger part is used to train the algorithm.
Due to the labels, 
the prediction of an algorithm can be compared with the label in order to determine whether the prediction was correct or not.
After training,
the remaining part is used for testing. 
In this phase,
the algorithm isn't adjusted anymore.
Still,
the predictions are compared to the labels in order to determine metrics that describe the quality of an outlier detection algorithm.
\\ \par
Outlier detection can be seen as a binary classifier:
An instance is either normal or anomalous.
There are several metrics available to determine the performance of a binary classifier.
For intrusion detection in industrial,
but also in home and office networks,
not only the number of detected attacks is relevant.
Due to the high amount of traffic, 
false positives have severe effects.
For one,
they need a lot of time to investigate.
Furthermore, 
they can,
on a psychological level,
have administrators become careless in cases of alarms as they expect them to be false positives,
the so called \textit{alarm fatigue}~\cite{Bliss.1995}.
Finally, 
the amount of normal traffic in networks usually outnumbers the amount of malicious traffic by magnitudes.
That means wrongly classifying 0.1\% of malicious and of normal traffic still results in vastly different numbers of alarms.
In this work,
we used two metrics to describe the performance of the algorithms:
The \textit{accuracy}~\cite{Olson.2008},
as well as the \textit{f-measure}~\cite{Rijsbergen.1979}.
The \textit{f-measure},
or \textit{F1-score},
is calculated as described in equation~\ref{eq:f-measure}.

\begin{equation}
\label{eq:f-measure}
F_{1} = 2 \cdot \dfrac{precision \cdot recall}{precision + recall}
\end{equation}

\begin{equation}
\label{eq:precision}
precision = \dfrac{t_{p}}{t_{p}+f_{p}}
\end{equation}

\begin{equation}
\label{eq:recall}
recall = \dfrac{t_{p}}{t_{p}+f_{n}}
\end{equation}

$t$ stands for a correct classification of the algorithm,
$f$ for an incorrect one.
An index $p$ indicates that the algorithm classified it as positive,
an index of $n$ indicates a classification as negative.
The \textit{F1-score} provides information about the relation of \textit{precision} and \textit{recall},
as defined in equations~\ref{eq:precision} and~\ref{eq:recall}.
\textit{Precision} and \textit{recall} describe the relation of all true positive classifications to all that have been classified as positive,
respectively to all events that are positive.
If both values are perfect,
the \textit{F1-score} amounts to one;
at worst,
it reaches 0.
The \textit{accuracy} is calculated according to equation~\ref{eq:accuracy}.

\begin{equation}
\label{eq:accuracy}
accuracy = \dfrac{t_{p} + t_{n}}{t_{p}+f_{p} + t_{n}+f_{n}}
\end{equation}

\textit{Accuracy} gives information about the relation of correct classifications in relation to all classifications.
First,
a naive approach to find outliers is described in subsection~\ref{ssec:naive}.
In subsections~\ref{ssec:svm} to~\ref{ssec:clustering},
the algorithms are applied to the three data sets \textit{DS1},
\textit{DS2} and \textit{DS3} and their performance is evaluated with the given metrics.
The results are then discussed in section~\ref{sec:res}.

\subsection{Naive Approach}
\label{ssec:naive}
In some data sets,
exploratory data analysis can lead to the discovery of singular or a group of features that can be used to distinguish between normal and anomalous data.
In \textit{DS1},
there are three features of the derived features as explained in table~\ref{tab:derived_features} capable of splitting the data set perfectly.
These features with the according values are listed in table~\ref{tab:naive_ds1}.
\begin{table}[!h]
\renewcommand{\arraystretch}{1.3}
\centering
\scriptsize
\caption{Features Capable of Perfectly Splitting \textit{DS1}}
\label{tab:naive_ds1}
\begin{tabular}{ l c c }
\toprule
\textbf{Feature} & \textbf{Norm. Values [Packets/s]}  & \textbf{Anom. Values [Packets/s]} \\
\textit{packets\_per\_sec} & 162, 164 & 3-9, 41 \\
\textit{max\_packets} & 72 & 208, 235, 401 \\
\textit{mean\_packets} &\numprint{58.28}, \numprint{58.48} & 105-150 \\
\bottomrule
\end{tabular}
\end{table}
This makes the application of machine learning algorithms obsolete,
all of the applied algorithms have to compete against a perfect score.
It is noteworthy,
however,
that these are derived and not basic features.
So at least a thorough understanding and sound feature engineering are necessary in order to be able to make sense of the data.
\\ \par
For \textit{DS2} and \textit{DS3},
no such features exist.
\textit{DS2} is too large with too few anomalous instances,
so that each feature of an anomalous event takes the same value on at least one other normal event.
\textit{DS3} is even more difficult,
as several data traces are mixed.
This leads to more heterogeneous feature distributions,
making it impossible to classify it by exploratory data analysis.

\subsection{\textit{Support Vector Machines}}
\label{ssec:svm}

\ac{svm} were first introduced by \textit{Boser et al.} in 1992~\cite{Boser.1992}.
The idea is to create a divider between two groups in such a way that each instance has the most possible distance from the divider.
This is called a \textit{large margin classifier}.
In \ac{svm},
data points are described by tuples as shown in equation~\ref{eq:svm_data}~\cite{Cortes.1995}.

\begin{equation}
\label{eq:svm_data}
\begin{split}
(x_{i}, y_{i}), i = 1, ..., m, y \in \{-1, 1\}
\end{split}
\end{equation}

$x$ is a vector describing a data point in an $n$-dimensional feature space.
$y$ describes the attribution to one of the two classes.
$m$ is the number of data points.
After training data,
the attribution is performed by the signum-function,
as shown in equation~\ref{eq:svm_attr}.
$w$ is the normal vector of the separator hyperplane,
$b$ is the offset from the hyperplane.

\begin{equation}
\label{eq:svm_attr}
\begin{split}
y_{i} = sgn(w, x_{i} - b)
\end{split}
\end{equation}

Generally,
when applying \ac{svm},
there are two different cases:
Either the set of instances can or can not be divided by a linear geometric figure.
If no linear division of the set of instances is possible,
the so-called \textit{kernel trick} is applied~\cite{Cortes.1995}.
In using the  \textit{kernel trick},
the input space is mapped non-linearly into a higher dimensional feature space,
where the algorithm can create a linear divider.
In this work,
the \textit{e1071}-library~\cite{e1071.2017} of the \textit{R} programming language~\cite{R.} has been used with a linear kernel.

\paragraph{DS1:}
\ac{svm} performs exceedingly well with this data set.
The relation of true and predicted labels can be found in table~\ref{tab:svm_ds1}.
\newcolumntype{g}{>{\columncolor{Gray}}l}
\begin{table}[!h]
\renewcommand{\arraystretch}{1.3}
\centering
\scriptsize
\caption{Predictions and Correct Labels of \textit{DS1} by Using \ac{svm}}
\label{tab:svm_ds1}
\begin{tabular}{ l  c  c  }
\toprule
\textbf{Label\textbackslash Prediction} & \textbf{Normal} & \textbf{Anomalous} \\
 \textbf{Normal} & \numprint{1097} & 0 \\
\textbf{Anomalous} & 0 & 22 \\
\bottomrule
\end{tabular}
\end{table}
\ac{svm} is capable of predicting each instance of the test data set correctly,
leading to an \textit{accuracy},
as well as \textit{F1-score} of 1,
as shown in line 1 of  table~\ref{tab:svm_ds_res}.
\begin{table}[!h]
\renewcommand{\arraystretch}{1.3}
\centering
\scriptsize
\caption{\textit{Accuracy} and \textit{F1-score} of \ac{svm}}
\label{tab:svm_ds_res}
\begin{tabular}{l c c}
\toprule
\textbf{Dataset} & \textbf{\textit{Accuracy}} & \textbf{\textit{F1-score}} \\
\textbf{\textit{DS1}} & \numprint{1.0} & \numprint{1.0} \\
\textbf{\textit{DS2}} & \numprint{1.0} & \numprint{1.0} \\
\textbf{\textit{DS3}} & \numprint{0.999936} &\numprint{0.999968} \\
\bottomrule
\end{tabular}
\end{table}

\paragraph{DS2:}
\ac{svm} performs exceedingly well with this data set as well.
The relation of true and predicted labels can be found in table~\ref{tab:svm_ds2}.
\begin{table}[!h]
\renewcommand{\arraystretch}{1.3}
\centering
\scriptsize
\caption{Predictions and Correct Labels of \textit{DS2} by Using \ac{svm}}
\label{tab:svm_ds2}
\begin{tabular}{l c c}
\toprule
\textbf{Label\textbackslash Prediction} & \textbf{Normal} & \textbf{Anomalous} \\
\textbf{Normal} & \numprint{3364} & 0 \\
\textbf{Anomalous} & 0 & 3 \\
\bottomrule
\end{tabular}
\end{table}
\ac{svm} is capable of predicting each instance of the test data set correctly,
leading to an \textit{accuracy},
as well as \textit{F1-score} of 1,
as shown in line 2 of table~\ref{tab:svm_ds_res}.

\paragraph{DS3:}
For this data set,
\ac{svm} still performs relatively well.
The relation of true and predicted labels can be found in table~\ref{tab:svm_ds2}.
\begin{table}[!h]
\renewcommand{\arraystretch}{1.3}
\centering
\scriptsize
\caption{Predictions and Correct Labels of \textit{DS2} by Using \ac{svm}}
\label{tab:svm_ds2}
\begin{tabular}{l c c}
\toprule
\textbf{Label\textbackslash Prediction} & \textbf{Normal} & \textbf{Anomalous} \\
\textbf{Normal} & \numprint{109702} & 4 \\
\textbf{Anomalous} & 3 & 63 \\
\bottomrule
\end{tabular}
\end{table}
\ac{svm} is capable of predicting most instances of the data set correctly,
indicated by \textit{accuracy} and \textit{F1-score} as shown in line 3 of table~\ref{tab:svm_ds_res}.

\subsection{\textit{Random Forrest}}
\label{ssec:random_forest}

A collection of \textit{Decision Trees} is called a \textit{Random Forest}~\cite{Breiman.2001}.
It consists of a \textit{root} node,
internal nodes,
so-called \textit{split} nodes and \textit{leaf} nodes.
Each  \textit{leaf} node corresponds to a class predicted by the \textit{Random Forest}.
All \textit{Decision Trees} have been grown during a training phase.
The final decision is made by a majority voting.
\textit{Random Forests} are robust to noise and overfitting,
a common problem in machine learning.
It happens when an algorithm puts too much importance on singular features so that instances of one class with less expressive characteristic of this feature are no longer classified correctly.
Furthermore,
they converge quickly.
In this work,
2 000 trees were used,
created by \textit{rpart}~\cite{rpart.2018} and \textit{randomForest}~\cite{randomForest.2015} in \textit{R}~\cite{R.}.

\paragraph{DS1:}
The \textit{Random Forest} algorithm performs well on this data set,
as shown in table~\ref{tab:rf_ds1}.
It  reaches a perfect score,
as depicted in line 1 of table~\ref{tab:rf_ds_res}.
\begin{table}[!h]
\renewcommand{\arraystretch}{1.3}
\centering
\scriptsize
\caption{Predictions and Correct Labels of \textit{DS1} by Using \textit{Random Forest}}
\label{tab:rf_ds1}
\begin{tabular}{l c c}
\toprule
\textbf{Label\textbackslash Prediction} & \textbf{Normal} & \textbf{Anomalous} \\
\textbf{Normal} & \numprint{973} & 0 \\
\textbf{Anomalous} & 0 & 23 \\
\bottomrule
\end{tabular}
\end{table}
\begin{table}[!h]
\renewcommand{\arraystretch}{1.3}
\centering
\scriptsize
\caption{\textit{Accuracy} and \textit{F1-score} of \textit{Random Forest}}
\label{tab:rf_ds_res}
\begin{tabular}{l c c}
\toprule
\textbf{Dataset} & \textbf{\textit{Accuracy}} & \textbf{\textit{F1-score}} \\
\textbf{\textit{DS1}} & \numprint{1.0} & \numprint{1.0} \\
\textbf{\textit{DS2}} & \numprint{0.999701} & \numprint{0.999851} \\
\textbf{\textit{DS3}} & \numprint{0.999973} & \numprint{0.999986} \\
\bottomrule
\end{tabular}
\end{table}

\paragraph{DS2:}
For this data set,
the \textit{Random Forest} algorithm obtains the worst results of all data sets.
The results are shown in table~\ref{tab:rf_ds2}.
But since the number of anomalous instances in comparison to the size of the data set is tiny,
the relatively poor results,
shown in line 2 of table~\ref{tab:rf_ds_res},
derive from the metrics and the weighting of its factors.
\begin{table}[!h]
\renewcommand{\arraystretch}{1.3}
\centering
\scriptsize
\caption{Predictions and Correct Labels of \textit{DS2} by Using \textit{Random Forest}}
\label{tab:rf_ds2}
\begin{tabular}{l c c }
\toprule
\textbf{Label\textbackslash Prediction} & \textbf{Normal} & \textbf{Anomalous} \\
\textbf{Normal} & \numprint{3347} & 1 \\
\textbf{Anomalous} & 0 & 2 \\
\bottomrule
\end{tabular}
\end{table}

\paragraph{DS3:}
In this data set,
the \textit{Random Forest} algorithm performs very well again,
the results are shown in table~\ref{tab:rf_ds3}.
No false negatives occur.
It even outperforms the \textit{SVM},
as shown in line 3 of table~\ref{tab:rf_ds_res}.
\begin{table}[!h]
\renewcommand{\arraystretch}{1.3}
\centering
\scriptsize
\caption{Predictions and Correct Labels of \textit{DS3} by Using \textit{Random Forest}}
\label{tab:rf_ds3}
\begin{tabular}{l c c }
\toprule
\textbf{Label\textbackslash Prediction} & \textbf{Normal} & \textbf{Anomalous} \\
\textbf{Normal} & \numprint{109710} & 3 \\
\textbf{Anomalous} & 0 & 59 \\
\bottomrule
\end{tabular}
\end{table}

\subsection{\textit{k-nearest Neighbour}}
\label{ssec:knn}
This algorithm is a non-parametric classification and regression algorithm~\cite{Altman.1992}.
In classification,
the affiliation of an event to a group is calculated by determining the set of the $k$ nearest neighbours,
commonly by calculating the Euclidean distance in an $n$-dimensional feature space as shown in equation~\ref{eq:euclidean}.
The event under evaluation is classified as part of the group with which it has the most common neighbours among its $k$ nearest ones.

\begin{equation}
\label{eq:euclidean}
D = \sqrt{\sum_{i=1}^{n}(x_{i}-w_{i})^{2}}
\end{equation}

As discussed before,
this is the only algorithm where the 80\%/20\% split led to an increased cross-validation result.

\paragraph{DS1:}
The performance of the \textit{k-nearest neighbour} algorithm on this data set is poor.
The relatively high false positive rate,
as shown in table~\ref{tab:knn_ds1},
leads to bad overall performance,
as shown in line 1 of table~\ref{tab:knn_ds_res}.
\begin{table}[!h]
\renewcommand{\arraystretch}{1.3}
\centering
\scriptsize
\caption{Predictions and Correct Labels of \textit{DS1} by Using \textit{k-nearest Neighbour}}
\label{tab:knn_ds1}
\begin{tabular}{l c c }
\toprule
\textbf{Label\textbackslash Prediction} & \textbf{Normal} & \textbf{Anomalous} \\
\textbf{Normal} & \numprint{678} & 0 \\
\textbf{Anomalous} & 2 & 9 \\
\bottomrule
\end{tabular}
\end{table}
\begin{table}[!h]
\renewcommand{\arraystretch}{1.3}
\centering
\scriptsize
\caption{\textit{Accuracy} and \textit{F1-score} of \textit{k-nearest Neighbour}}
\label{tab:knn_ds_res}
\begin{tabular}{l c c }
\toprule
\textbf{Dataset} & \textbf{\textit{Accuracy}} & \textbf{\textit{F1-score}} \\
\textbf{\textit{DS1}} & \numprint{0.997097} & \numprint{0.998527} \\
\textbf{\textit{DS2}} & \numprint{0.999118} & \numprint{0.999559} \\
\textbf{\textit{DS3}} & \numprint{0.999412} & \numprint{0.999706} \\
\bottomrule
\end{tabular}
\end{table}

\paragraph{DS2:}
Even though the \textit{k-nearest neighbour} algorithm performs better on this data set,
it is still not satisfying.
The algorithm classifies any event as normal,
as shown in table~\ref{tab:knn_ds2}.
the small amount of anomalous events still leads to a medium performance evaluation,
as shown in line 2 of table~\ref{tab:knn_ds_res}.
\begin{table}[!h]
\renewcommand{\arraystretch}{1.3}
\centering
\scriptsize
\caption{Predictions and Correct Labels of \textit{DS2} by Using \textit{k-nearest Neighbour}}
\label{tab:knn_ds2}
\begin{tabular}{l c c }
\toprule
\textbf{Label\textbackslash Prediction} & \textbf{Normal} & \textbf{Anomalous} \\
\textbf{Normal} & \numprint{2265} & 0 \\
\textbf{Anomalous} & 2 & 0 \\
\bottomrule
\end{tabular}
\end{table}

\paragraph{DS3:}
As in applying the \textit{k-nearest neighbour} algorithm to \textit{DS2},
it classifies each instance of \textit{DS3} as normal as well.
This is shown in table~\ref{tab:knn_ds3}.
The according metrics can be found in line 3 of table~\ref{tab:knn_ds_res}.
\begin{table}[!h]
\renewcommand{\arraystretch}{1.3}
\centering
\scriptsize
\caption{Predictions and Correct Labels of \textit{DS3} by Using \textit{k-nearest Neighbour}}
\label{tab:knn_ds3}
\begin{tabular}{l c c }
\toprule
\textbf{Label\textbackslash Prediction} & \textbf{Normal} & \textbf{Anomalous} \\
\textbf{Normal} & \numprint{73140} & 0 \\
\textbf{Anomalous} & 43 & 0 \\
\bottomrule
\end{tabular}
\end{table}

\subsection{\textit{k-means Clustering}}
\label{ssec:clustering}

In \textit{k-means clustering}~\cite{Alsabti.1997},
the probability of an object belonging to a group is calculated.
This probability is commonly calculated as the Euclidean distance,
as introduced in equation~\ref{eq:euclidean},
of a point in an $n$-dimensional feature space from the center of a cluster.
In applying the \textit{k-means}-algorithm
those distances are minimized with an error function as shown in equation~\ref{eq:k-means}.

\begin{equation}
\label{eq:k-means}
\begin{split}
E = \sum\limits_{j=1}^{k} \sum\limits_{i_{l} \in C_{j}} | i_{l} - w_{j} |^{2} \\
j \in \{1, ..., k\}, l \in \{1, ..., n\}
\end{split}
\end{equation}

$k$ is the number of clusters, 
that needs to be defined a priori.
In this work,
two clusters were used,
one to describe normal,
the other to describe anomalous behaviour.
$n$ is the number of events or elements in the feature space and $C$ is the cluster.
In contrast to the above algorithms,
there cannot be a comparison between label and prediction.
Instead,
each of the two clusters has to be given a label in order to determine the quality.
In this work,
clusters were chosen as if a users did not have labels to support decision making,
which is also the choice that minimises the error.
This means that the cluster containing the larger portion of elements is seen as the cluster with label ``normal''.
Furthermore,
\textit{k-means clustering} is the only algorithm considered in this work that is non-supervised,
meaning it does not need training.

The biggest advantage of non-supervised machine learning algorithms is omitting the need to find a valid training data set.
On the other hand,
if they are applied to unlabeled data, 
it is hard to determine their performance.

\paragraph{DS1:}
In applying \textit{k-means clustering} to this data set,
all normal events are grouped in one cluster.
Most of the anomalous events,
however, 
are clustered there as well,
as shown in table~\ref{tab:kmc_ds1}.
The according \textit{accuracy} and \textit{F1-score} can be found in line 1 of table~\ref{tab:kmc_ds_res}.
\begin{table}[!h]
\renewcommand{\arraystretch}{1.3}
\centering
\scriptsize
\caption{Predictions and Clusters of \textit{DS1} by Using \textit{k-means Clustering}}
\label{tab:kmc_ds1}
\begin{tabular}{l c c }
\toprule
\textbf{Cluster\textbackslash Label} & \textbf{Normal} & \textbf{Anomalous} \\
\textbf{Cluster 1} & 0 & 12 \\
\textbf{Cluster 2} & \numprint{3244} & 63 \\
\bottomrule
\end{tabular}
\end{table}
\begin{table}[!h]
\renewcommand{\arraystretch}{1.3}
\centering
\scriptsize
\caption{\textit{Accuracy} and \textit{F1-score} of \textit{k-means Clustering}}
\label{tab:kmc_ds_res}
\begin{tabular}{l c c }
\toprule
\textbf{Dataset} & \textbf{\textit{Accuracy}} & \textbf{\textit{F1-score}} \\
\textbf{\textit{DS1}} & \numprint{0.981018} & \numprint{0.990383} \\
\textbf{\textit{DS2}} & \numprint{0.556242} & \numprint{0.714853} \\
\textbf{\textit{DS3}} & \numprint{0.633624} & \numprint{0.775728} \\
\bottomrule
\end{tabular}
\end{table}

\paragraph{DS2:}
The \textit{k-means clustering}-algorithm distributes the ``normal''-labeled events in both clusters,
in a comparable amount (about \numprint{5000} vs. \numprint{6200}) as shown in table~\ref{tab:kmc_ds2}.
This leads to significantly reduced performance metrics,
listed in line 2 of table~\ref{tab:kmc_ds_res}.
Furthermore,
all anomalous events are grouped in the larger cluster,
classifying them as normal.
\begin{table}[!h]
\renewcommand{\arraystretch}{1.3}
\centering
\scriptsize
\caption{Predictions and Clusters of \textit{DS2} by Using \textit{k-means Clustering}}
\label{tab:kmc_ds2}
\begin{tabular}{l c c }
\toprule
\textbf{Clusters\textbackslash Label} & \textbf{Normal} & \textbf{Anomalous} \\
\textbf{Clusters 1} & \numprint{4945} & 0 \\
\textbf{Clusters 2} & \numprint{6211} & 10 \\
\bottomrule
\end{tabular}
\end{table}

\paragraph{DS3:}
As in applying \textit{k-means clustering} to \textit{DS2},
all events labeled ``anomalous'' in this data set are grouped in the same cluster as most of the events labeled ``normal''.
This effect is depicted in table~\ref{tab:kmc_ds3}.
Since there are about \numprint{2000} times as many normal events as anomalous,
the performance metrics are slightly improved in comparison to the above use case,
as listed in line 3 of table~\ref{tab:kmc_ds3}.
\begin{table}[!h]
\renewcommand{\arraystretch}{1.3}
\centering
\scriptsize
\caption{Predictions and Clusters of \textit{DS3} by Using \textit{k-means Clustering}}
\label{tab:kmc_ds3}
\begin{tabular}{l c c }
\toprule
\textbf{Clusters\textbackslash Label} & \textbf{Normal} & \textbf{Anomalous} \\
\textbf{Clusters 1} & \numprint{231847} & \numprint{206} \\
\textbf{Clusters 2} & \numprint{133853} & 0 \\
\bottomrule
\end{tabular}
\end{table}

\section{Results and Discussion}
\label{sec:res}
\textit{DS1} can be seen as a sort of necessary condition:
since it is perfectly separatable based on a three derived features,
as described in subsection~\ref{ssec:naive},
the algorithms should lead to a perfect result as well.
Only \ac{svm} and \textit{Random Forest} did so.
Both of them performed very well on the other two data sets as well,
\ac{svm} outperformed \textit{Random Forest} on \textit{DS2},
and vice versa on \textit{DS3}. 
\textit{k-nearest neighbour} and \textit{k-means clustering} performed significantly worse.
In machine learning,
\textit{F1-scores} and \textit{accuracy} scores of around \numprint{0.9999} are usually required in order to consider the performance of a given algorithm good.
While \textit{k-nearest neighbour} is sometimes close to these values,
\textit{k-means clustering} leads to results far from satisfying.
Maybe,
optimizing the number of clusters,
e.g. by calculating and maximising the \textit{silhouette} coefficients~\cite{Rousseeuw.1987},
would improve the performance.
\\ \par
In their work,
\textit{Lemay and Fernandez} state that the regularity of their traffic would make it easy for machine learning-based anomaly detection algorithms to find the attacks.
This is especially true for \textit{DS1}.
They also provide data sets covert channel attacks that are more subtle~\cite{Lemay.2016}.
To increase the difficulty for the algorithms,
and to mimic the changing nature of real industrial applications,
we mixed several data sets in \textit{DS3}.
Still,
\textit{Random Forest} and \ac{svm} were able to find an impressive number of attacks.
\\ \par
Furthermore,
it should be noted that all of the features used for detection are ethernet- and \ac{tcp}/\ac{ip}-based.
The \textit{Modbus} protocol-based characteristics did not have any direct influence on the detection mechanisms.
However,
the regularity and the structure of the traffic differs significantly from home- and office-based network traffic.
This means that industrial traffic is different in character and thus different in detecting by algorithm,
even if no protocol-specific attributes are employed.

\section{Conclusion and Outlook}
\label{sec:conclusion}
In this work,
it is shown that some machine learning-based anomaly detection algorithms,
in this case namely \ac{svm} and \textit{Random Forest},
perform well in detecting network traffic anomalies in industrial networks.
Since both of them are supervised methods,
however,
training data is needed.
This data can be provided by simulators,
as the one of \textit{Lemay and Fernandez},
that was analysed in this work.
The difficulty lies in generating sound, valid data that matches the industrial environment in which the anomaly detection algorithm shall be applied.
\\ \par
There are several possibilities for extension of the presented methods.
Data from different sources can be gathered,
combined and used to enhance the results~\cite{Duque_Anton.2017b}.
The introduction of context information into the anomaly detection process is promising and capable of increasing the performance~\cite{Duque_Anton.2017c}.
Furthermore,
the employment of \textit{deception technologies} as sensors for anomaly detection could be used to enhance the insight about malicious behaviour~\cite{Fraunholz.2017a}.
\\ \par
One of the most prevalent necessities is the generation of data with attacks that are specific to industrial applications in general,
and especially to \textit{Modbus}.
The analysis performed in this work merely employs network-based features that,
in the same form,
exist in home and office appliances.
The only major difference is the timing pattern that is strongly correlated to attacks.


\section*{Acknowledgments}
This work has been supported by the Federal Ministry of Education and Research of the Federal Republic of Germany (Foerderkennzeichen KIS4ITS0001, IUNO).
The authors alone are responsible for the content of the paper.

\bibliographystyle{ACM-Reference-Format}
\bibliography{paper}


\begin{thebibliography}{00}


\ifx \showCODEN    \undefined \def \showCODEN     #1{\unskip}     \fi
\ifx \showDOI      \undefined \def \showDOI       #1{#1}\fi
\ifx \showISBNx    \undefined \def \showISBNx     #1{\unskip}     \fi
\ifx \showISBNxiii \undefined \def \showISBNxiii  #1{\unskip}     \fi
\ifx \showISSN     \undefined \def \showISSN      #1{\unskip}     \fi
\ifx \showLCCN     \undefined \def \showLCCN      #1{\unskip}     \fi
\ifx \shownote     \undefined \def \shownote      #1{#1}          \fi
\ifx \showarticletitle \undefined \def \showarticletitle #1{#1}   \fi
\ifx \showURL      \undefined \def \showURL       {\relax}        \fi
\providecommand\bibfield[2]{#2}
\providecommand\bibinfo[2]{#2}
\providecommand\natexlab[1]{#1}
\providecommand\showeprint[2][]{arXiv:#2}

\bibitem[\protect\citeauthoryear{??}{rpa}{2018}]%
        {rpart.2018}
 \bibinfo{year}{2018}\natexlab{}.
\newblock \bibinfo{booktitle}{{\em {Package rpart}}}.
\newblock


\bibitem[\protect\citeauthoryear{Alsabti, Ranka, and Singh}{Alsabti
  et~al\mbox{.}}{1997}]%
        {Alsabti.1997}
\bibfield{author}{\bibinfo{person}{Khaled Alsabti}, \bibinfo{person}{Sanjay
  Ranka}, {and} \bibinfo{person}{Vineet Singh}.}
  \bibinfo{year}{1997}\natexlab{}.
\newblock \showarticletitle{{An efficient k-means clustering algorithm}}.
\newblock \bibinfo{journal}{{\em Electrical Engineering and Computer
  Science\/}} (\bibinfo{date}{January} \bibinfo{year}{1997}).
\newblock


\bibitem[\protect\citeauthoryear{Altman}{Altman}{1992}]%
        {Altman.1992}
\bibfield{author}{\bibinfo{person}{N.~S. Altman}.}
  \bibinfo{year}{1992}\natexlab{}.
\newblock \showarticletitle{{An Introduction to Kernel and Nearest-Neighbor
  Nonparametric Regression}}.
\newblock \bibinfo{journal}{{\em The American Statistician\/}}
  \bibinfo{volume}{46}, \bibinfo{number}{3} (\bibinfo{date}{August}
  \bibinfo{year}{1992}), \bibinfo{pages}{175--185}.
\newblock


\bibitem[\protect\citeauthoryear{Bhuyan, Bhattacharyya, and Kalita}{Bhuyan
  et~al\mbox{.}}{2014}]%
        {Bhuyan.2014}
\bibfield{author}{\bibinfo{person}{Monowar~H. Bhuyan}, \bibinfo{person}{D.~K.
  Bhattacharyya}, {and} \bibinfo{person}{J.~K. Kalita}.}
  \bibinfo{year}{2014}\natexlab{}.
\newblock \showarticletitle{{Network Anomaly Detection: Methods, Systems and
  Tools}}.
\newblock \bibinfo{journal}{{\em IEEE Communications Surveys {\&} Tutorials\/}}
  \bibinfo{volume}{16}, \bibinfo{number}{1} (\bibinfo{year}{2014}),
  \bibinfo{pages}{303--336}.
\newblock
\showDOI{%
\url{https://doi.org/10.1109/SURV.2013.052213.00046}}


\bibitem[\protect\citeauthoryear{Bliss, Gilson, and Deaton}{Bliss
  et~al\mbox{.}}{1995}]%
        {Bliss.1995}
\bibfield{author}{\bibinfo{person}{James Bliss}, \bibinfo{person}{Richard~D.
  Gilson}, {and} \bibinfo{person}{John~E. Deaton}.}
  \bibinfo{year}{1995}\natexlab{}.
\newblock \showarticletitle{{Human probability matching behaviour in response
  to alarms of varying reliability}}.
\newblock \bibinfo{journal}{{\em Ergonomics\/}} \bibinfo{volume}{38},
  \bibinfo{number}{11} (\bibinfo{date}{December} \bibinfo{year}{1995}).
\newblock
\showDOI{%
\url{https://doi.org/10.1080/00140139508925269}}


\bibitem[\protect\citeauthoryear{Boser, Guyon, and Vapnik}{Boser
  et~al\mbox{.}}{1992}]%
        {Boser.1992}
\bibfield{author}{\bibinfo{person}{Bernhard~E. Boser},
  \bibinfo{person}{Isabelle~M. Guyon}, {and} \bibinfo{person}{Vladimir~N.
  Vapnik}.} \bibinfo{year}{1992}\natexlab{}.
\newblock \showarticletitle{{A Training Algorithm for Optimal Margin
  Classifiers}}. In \bibinfo{booktitle}{{\em Proceedings of the Fifth Annual
  Workshop on Computational Learning Theory}} {\em (\bibinfo{series}{COLT
  '92})}. \bibinfo{publisher}{ACM}, \bibinfo{address}{New York, NY, USA},
  \bibinfo{pages}{144--152}.
\newblock
\showISBNx{0-89791-497-X}
\showDOI{%
\url{https://doi.org/10.1145/130385.130401}}


\bibitem[\protect\citeauthoryear{Breiman}{Breiman}{2001}]%
        {Breiman.2001}
\bibfield{author}{\bibinfo{person}{Leo Breiman}.}
  \bibinfo{year}{2001}\natexlab{}.
\newblock \showarticletitle{{Random Forests}}.
\newblock \bibinfo{journal}{{\em Machine Learning\/}} \bibinfo{volume}{45},
  \bibinfo{number}{1} (\bibinfo{date}{Octoober} \bibinfo{year}{2001}),
  \bibinfo{pages}{5--32}.
\newblock
\showDOI{%
\url{https://doi.org/10.1023/A:1010933404324}}


\bibitem[\protect\citeauthoryear{Cortes and Vapnik}{Cortes and Vapnik}{1995}]%
        {Cortes.1995}
\bibfield{author}{\bibinfo{person}{Corinna Cortes} {and}
  \bibinfo{person}{Vladimir Vapnik}.} \bibinfo{year}{1995}\natexlab{}.
\newblock \showarticletitle{{Support-Vector Networks}}.
\newblock \bibinfo{journal}{{\em Machine Learning\/}} \bibinfo{volume}{20},
  \bibinfo{number}{3} (\bibinfo{date}{September} \bibinfo{year}{1995}),
  \bibinfo{pages}{273--297}.
\newblock
\showISSN{0885-6125}
\showDOI{%
\url{https://doi.org/10.1023/A:1022627411411}}


\bibitem[\protect\citeauthoryear{Denning}{Denning}{1987}]%
        {Denning.1987}
\bibfield{author}{\bibinfo{person}{Dorothy~E. Denning}.}
  \bibinfo{year}{1987}\natexlab{}.
\newblock \showarticletitle{{An Intrusion-Detection Model}}.
\newblock \bibinfo{journal}{{\em IEEE Transactions on Software Engineering\/}}
  \bibinfo{volume}{SE-13}, \bibinfo{number}{2} (\bibinfo{date}{Feburary}
  \bibinfo{year}{1987}), \bibinfo{pages}{222--232}.
\newblock


\bibitem[\protect\citeauthoryear{Drury}{Drury}{2009}]%
        {Drury.2009}
\bibfield{author}{\bibinfo{person}{Bill Drury}.}
  \bibinfo{year}{2009}\natexlab{}.
\newblock \bibinfo{booktitle}{{\em {Control Techniques Drives and Controls
  Handbook}\/} (\bibinfo{edition}{2nd} ed.)}.
\newblock \bibinfo{publisher}{Institution of Engineering and Technology}.
\newblock
\showISBNx{978-1-84919-013-8}


\bibitem[\protect\citeauthoryear{Duque~Anton, Fraunholz, Lipps, Pohl,
  Zimmermann, and Schotten}{Duque~Anton et~al\mbox{.}}{2017a}]%
        {Duque_Anton.2017a}
\bibfield{author}{\bibinfo{person}{Simon Duque~Anton}, \bibinfo{person}{Daniel
  Fraunholz}, \bibinfo{person}{Christoph Lipps}, \bibinfo{person}{Frederic
  Pohl}, \bibinfo{person}{Marc Zimmermann}, {and} \bibinfo{person}{Hans~Dieter
  Schotten}.} \bibinfo{year}{2017}\natexlab{a}.
\newblock \showarticletitle{{Two Decades of SCADA Exploitation: A Brief
  History}}. In \bibinfo{booktitle}{{\em IEEE Conference on Applications,
  Information and Network Security (AINS). IEEE Conference on Applications,
  Information and Network Security (AINS-2017), November 13-14, Miri, Sarawak,
  Malaysia}}. IEEE Computer Science Chapter Malaysia, \bibinfo{publisher}{IEEE
  Press}.
\newblock
\showDOI{%
\url{https://doi.org/10.1109/AINS.2017.8270432}}


\bibitem[\protect\citeauthoryear{Duque~Anton, Fraunholz, Teuber, and
  Schotten}{Duque~Anton et~al\mbox{.}}{2017b}]%
        {Duque_Anton.2017c}
\bibfield{author}{\bibinfo{person}{Simon Duque~Anton}, \bibinfo{person}{Daniel
  Fraunholz}, \bibinfo{person}{Stephan Teuber}, {and}
  \bibinfo{person}{Hans~Dieter Schotten}.} \bibinfo{year}{2017}\natexlab{b}.
\newblock \showarticletitle{{A Question of Context: Enhancing Intrusion
  Detection by Providing Context Information}}. In \bibinfo{booktitle}{{\em
  13th Conference of Telecommunication, Media and Internet Techno-Economics
  (CTTE-17)}}. Aalborg University Copenhagen, \bibinfo{publisher}{IEEE}.
\newblock


\bibitem[\protect\citeauthoryear{Duque~Anton, Fraunholz, Zemitis, Pohl, and
  Schotten}{Duque~Anton et~al\mbox{.}}{2017c}]%
        {Duque_Anton.2017b}
\bibfield{author}{\bibinfo{person}{Simon Duque~Anton}, \bibinfo{person}{Daniel
  Fraunholz}, \bibinfo{person}{Janis Zemitis}, \bibinfo{person}{Frederic Pohl},
  {and} \bibinfo{person}{Hans~Dieter Schotten}.}
  \bibinfo{year}{2017}\natexlab{c}.
\newblock \showarticletitle{{Highly Scalable and Flexible Model for Effective
  Aggregation of Context-based Data in Generic IIoT Scenarios}}. In
  \bibinfo{booktitle}{{\em 9th Central European Workshop on Services and their
  Composition. Central European Workshop on Services and their Composition
  (ZEUS-2017), February 13-14, Lugano, Switzerland}},
  \bibfield{editor}{\bibinfo{person}{Oliver Kopp}, \bibinfo{person}{Jörg
  Lenhard}, {and} \bibinfo{person}{Cesare Pautasso}} (Eds.).
  \bibinfo{publisher}{CEUR Workshop Proceedings}, \bibinfo{pages}{51--58}.
\newblock


\bibitem[\protect\citeauthoryear{{EtherCAT Technology Group}}{{EtherCAT
  Technology Group}}{1991}]%
        {ethercat.2018}
\bibfield{author}{\bibinfo{person}{{EtherCAT Technology Group}}.}
  \bibinfo{year}{1991}\natexlab{}.
\newblock \bibinfo{title}{{EtherCAT - The Ethernet Fieldbus}}.
\newblock   (\bibinfo{year}{1991}).
\newblock
\showURL{%
\url{https://www.ethercat.org/default.htm}}


\bibitem[\protect\citeauthoryear{Foundation}{Foundation}{[n. d.]}]%
        {Python.}
\bibfield{author}{\bibinfo{person}{Python~Software Foundation}.}
  \bibinfo{year}{[n. d.]}\natexlab{}.
\newblock \bibinfo{title}{{Python}}.
\newblock   (\bibinfo{year}{[n. d.]}).
\newblock
\showURL{%
\url{https://www.python.org/}}


\bibitem[\protect\citeauthoryear{Fraunholz, Krohmer, Duque~Anton, and
  Schotten}{Fraunholz et~al\mbox{.}}{2017a}]%
        {Fraunholz.2017}
\bibfield{author}{\bibinfo{person}{Daniel Fraunholz}, \bibinfo{person}{Daniel
  Krohmer}, \bibinfo{person}{Simon Duque~Anton}, {and}
  \bibinfo{person}{Hans~Dieter Schotten}.} \bibinfo{year}{2017}\natexlab{a}.
\newblock \showarticletitle{Investigation of Cyber Crime Conducted by Abusing
  Weak or Default Passwords with a Medium Interaction Honeypot}. In
  \bibinfo{booktitle}{{\em International Conference On Cyber Security And
  Protection Of Digital Services(Cyber Security-17)}}.
  \bibinfo{publisher}{IEEE}.
\newblock


\bibitem[\protect\citeauthoryear{Fraunholz, Zimmermann, and Schotten}{Fraunholz
  et~al\mbox{.}}{2017b}]%
        {Fraunholz.2017a}
\bibfield{author}{\bibinfo{person}{Daniel Fraunholz}, \bibinfo{person}{Marc
  Zimmermann}, {and} \bibinfo{person}{Hans~Dieter Schotten}.}
  \bibinfo{year}{2017}\natexlab{b}.
\newblock \showarticletitle{{Towards Deployment Strategies for Deception
  Systems}}.
\newblock \bibinfo{journal}{{\em Advances in Science, Technology and
  Engineering Systems Journal (ASTESJ)\/}}  \bibinfo{volume}{Special Issue on
  Recent Advances in Engineering Systems 2017} (\bibinfo{date}{July}
  \bibinfo{year}{2017}), \bibinfo{pages}{1272--1279}.
\newblock


\bibitem[\protect\citeauthoryear{Genge, Siaterlis, Fovino, and Masera}{Genge
  et~al\mbox{.}}{2012}]%
        {Genge.2012}
\bibfield{author}{\bibinfo{person}{Bela Genge}, \bibinfo{person}{Christos
  Siaterlis}, \bibinfo{person}{Igor~Nai Fovino}, {and} \bibinfo{person}{Marcelo
  Masera}.} \bibinfo{year}{2012}\natexlab{}.
\newblock \showarticletitle{{A Cyber-Physical Experimentation Environment for
  the Security Analysis of Networked Industrial Control Systems}}.
\newblock \bibinfo{journal}{{\em Computers {\&} Electrical Engineering\/}}
  \bibinfo{volume}{38}, \bibinfo{number}{5} (\bibinfo{date}{September}
  \bibinfo{year}{2012}), \bibinfo{pages}{1146--1161}.
\newblock
\showDOI{%
\url{https://doi.org/10.1016/j.compeleceng.2012.06.015}}


\bibitem[\protect\citeauthoryear{Hadeli, Schierholz, Braendle, and
  Tuduce}{Hadeli et~al\mbox{.}}{2009}]%
        {Hadeli.2009}
\bibfield{author}{\bibinfo{person}{Hadeli Hadeli}, \bibinfo{person}{Ragnar
  Schierholz}, \bibinfo{person}{Markus Braendle}, {and}
  \bibinfo{person}{Christian Tuduce}.} \bibinfo{year}{2009}\natexlab{}.
\newblock \showarticletitle{{Leveraging determinism in industrial control
  systems for advanced anomaly detection and reliable security configuration}}.
  In \bibinfo{booktitle}{{\em 2009 IEEE Conference on Emerging Technologies
  Factory Automation}}. \bibinfo{pages}{1--8}.
\newblock
\showISSN{1946-0740}
\showDOI{%
\url{https://doi.org/10.1109/ETFA.2009.5347134}}


\bibitem[\protect\citeauthoryear{Hadziosmanovic, Bolzoni, Hartel, and
  Etalle}{Hadziosmanovic et~al\mbox{.}}{2011}]%
        {Hadziosmanovic.2011}
\bibfield{author}{\bibinfo{person}{Dina Hadziosmanovic},
  \bibinfo{person}{Damiano Bolzoni}, \bibinfo{person}{Pieter Hartel}, {and}
  \bibinfo{person}{Sandro Etalle}.} \bibinfo{year}{2011}\natexlab{}.
\newblock \bibinfo{booktitle}{{\em {MELISSA: Towards Automated Detection of
  Undesirable User Actions in Critical Infrastructures}}}.
\newblock \bibinfo{publisher}{IEEE Computer Society}, \bibinfo{pages}{41--48}.
\newblock
\showISBNx{978-0-7695-4762-6}
\showDOI{%
\url{https://doi.org/10.1109/EC2ND.2011.10}}


\bibitem[\protect\citeauthoryear{Igure, Laughter, and Williams}{Igure
  et~al\mbox{.}}{2006}]%
        {Igure.2006}
\bibfield{author}{\bibinfo{person}{Vinay~M. Igure}, \bibinfo{person}{Sean~A.
  Laughter}, {and} \bibinfo{person}{Ronald~D. Williams}.}
  \bibinfo{year}{2006}\natexlab{}.
\newblock \showarticletitle{{Security issues in SCADA networks}}.
\newblock \bibinfo{journal}{{\em Computers {\&} Security\/}}
  \bibinfo{volume}{25} (\bibinfo{year}{2006}), \bibinfo{pages}{498--506}.
\newblock
\showDOI{%
\url{https://doi.org/10.1016/j.cose.2006.03.001}}


\bibitem[\protect\citeauthoryear{Igure, Laughter, and Williams}{Igure
  et~al\mbox{.}}{2009}]%
        {Garcia-Teodoro.2009}
\bibfield{author}{\bibinfo{person}{Vinay~M. Igure}, \bibinfo{person}{Sean~A.
  Laughter}, {and} \bibinfo{person}{Ronald~D. Williams}.}
  \bibinfo{year}{2009}\natexlab{}.
\newblock \showarticletitle{{Anomaly-based network intrusion detection:
  Techniques, systems and challenges}}.
\newblock \bibinfo{journal}{{\em Computers {\&} Security\/}}
  \bibinfo{number}{28} (\bibinfo{date}{February} \bibinfo{year}{2009}),
  \bibinfo{pages}{18--28}.
\newblock
\showDOI{%
\url{https://doi.org/10.1016/j.cose.2008.08.003}}


\bibitem[\protect\citeauthoryear{Kleinmann and Wool}{Kleinmann and
  Wool}{2014}]%
        {Kleinmann.2014}
\bibfield{author}{\bibinfo{person}{Amit Kleinmann} {and}
  \bibinfo{person}{Avishai Wool}.} \bibinfo{year}{2014}\natexlab{}.
\newblock \showarticletitle{{Accurate Modeling of the Siemens S7 SCADA Protocol
  for Intrusion Detection and Digital Forensics}}. In \bibinfo{booktitle}{{\em
  Journal of Digital Forensics, Security and Law}}, Vol.~\bibinfo{volume}{9}.
\newblock


\bibitem[\protect\citeauthoryear{Lemay and Fernandez}{Lemay and
  Fernandez}{2016}]%
        {Lemay.2016}
\bibfield{author}{\bibinfo{person}{Antoine Lemay} {and}
  \bibinfo{person}{Jose~M. Fernandez}.} \bibinfo{year}{2016}\natexlab{}.
\newblock \showarticletitle{{Providing SCADA Network Data Sets for Intrusion
  Detection Research}}. In \bibinfo{booktitle}{{\em 9th Workshop on Cyber
  Security Experimentation and Test ({CSET} 16)}}. \bibinfo{publisher}{{USENIX}
  Association}, \bibinfo{address}{Austin, TX}.
\newblock
\showURL{%
\url{https://www.usenix.org/conference/cset16/workshop-program/presentation/lemay}}


\bibitem[\protect\citeauthoryear{Mantere, Sailio, and Noponen}{Mantere
  et~al\mbox{.}}{2013}]%
        {Mantere.2013}
\bibfield{author}{\bibinfo{person}{Matti Mantere}, \bibinfo{person}{Mirko
  Sailio}, {and} \bibinfo{person}{Sami Noponen}.}
  \bibinfo{year}{2013}\natexlab{}.
\newblock \showarticletitle{{Network Traffic Features for Anomaly Detection in
  Specific Industrial Control System Network}}.
\newblock \bibinfo{journal}{{\em Future Internet\/}} \bibinfo{volume}{4},
  \bibinfo{number}{5} (\bibinfo{date}{September} \bibinfo{year}{2013}),
  \bibinfo{pages}{460--473}.
\newblock
\showDOI{%
\url{https://doi.org/10.3390/fi5040460}}


\bibitem[\protect\citeauthoryear{Mathworks}{Mathworks}{[n. d.]}]%
        {Simulink.}
\bibfield{author}{\bibinfo{person}{Mathworks}.} \bibinfo{year}{[n.
  d.]}\natexlab{}.
\newblock \bibinfo{title}{{Simulation and Model-Based Design}}.
\newblock   (\bibinfo{year}{[n. d.]}).
\newblock
\showURL{%
\url{https://www.mathworks.com/products/simulink.html}}


\bibitem[\protect\citeauthoryear{Meshram and Haas}{Meshram and Haas}{2016}]%
        {Meshram.2016}
\bibfield{author}{\bibinfo{person}{Ankush Meshram} {and}
  \bibinfo{person}{Christian Haas}.} \bibinfo{year}{2016}\natexlab{}.
\newblock \showarticletitle{{Anomaly Detection in Industrial Networks using
  Machine Learning: A Roadmap}}. In \bibinfo{booktitle}{{\em {Machine Learning
  for Cyber Physical Systems}}}. \bibinfo{pages}{65--72}.
\newblock


\bibitem[\protect\citeauthoryear{Modbus}{Modbus}{2012}]%
        {Modbus.2012}
\bibfield{author}{\bibinfo{person}{Modbus}.} \bibinfo{year}{2012}\natexlab{}.
\newblock \bibinfo{title}{MODBUS APPLICATION PROTOCOL SPECIFICATION V1.1b3}.
\newblock   (\bibinfo{year}{2012}).
\newblock
\showURL{%
\url{http://www.modbus.org/docs/Modbus_Application_Protocol_V1_1b3.pdf}}


\bibitem[\protect\citeauthoryear{Modbus-IDA}{Modbus-IDA}{2006}]%
        {ModbusIDA.2006}
\bibfield{author}{\bibinfo{person}{Modbus-IDA}.}
  \bibinfo{year}{2006}\natexlab{}.
\newblock \bibinfo{title}{MODBUS MESSAGING ON TCP/IP IMPLEMENTATION GUIDE
  V1.0b}.
\newblock   (\bibinfo{year}{2006}).
\newblock
\showURL{%
\url{http://www.modbus.org/docs/Modbus_Messaging_Implementation_Guide_V1_0b.pdf}}


\bibitem[\protect\citeauthoryear{{MODICON Inc.}}{{MODICON Inc.}}{1996}]%
        {MODICONInc..1996}
\bibfield{author}{\bibinfo{person}{{MODICON Inc.}}}
  \bibinfo{year}{1996}\natexlab{}.
\newblock   (\bibinfo{year}{1996}).
\newblock
\showURL{%
\url{http://www.modbus.org/docs/PI_MBUS_300.pdf}}


\bibitem[\protect\citeauthoryear{Morris and Gao}{Morris and Gao}{2014}]%
        {Morris.2014}
\bibfield{author}{\bibinfo{person}{Thomas Morris} {and} \bibinfo{person}{Wei
  Gao}.} \bibinfo{year}{2014}\natexlab{}.
\newblock \bibinfo{booktitle}{{\em {Industrial Control System Traffic Data Sets
  for Intrusion Detection Research}}}.
\newblock \bibinfo{publisher}{Springer Berlin Heidelberg},
  \bibinfo{address}{Berlin, Heidelberg}, \bibinfo{pages}{65--78}.
\newblock
\showISBNx{978-3-662-45355-1}
\showDOI{%
\url{https://doi.org/10.1007/978-3-662-45355-1_5}}


\bibitem[\protect\citeauthoryear{{Morris, Thomas}}{{Morris, Thomas}}{[n. d.]}]%
        {Morris.}
\bibfield{author}{\bibinfo{person}{{Morris, Thomas}}.} \bibinfo{year}{[n.
  d.]}\natexlab{}.
\newblock \bibinfo{title}{{Industrial Control System (ICS) Cyber Attack
  Datasets}}.
\newblock   (\bibinfo{year}{[n. d.]}).
\newblock
\showURL{%
\url{https://sites.google.com/a/uah.edu/tommy-morris-uah/ics-data-sets}}


\bibitem[\protect\citeauthoryear{Olson and Dursun}{Olson and Dursun}{[n. d.]}]%
        {Olson.2008}
\bibfield{author}{\bibinfo{person}{David~L. Olson} {and} \bibinfo{person}{Delen
  Dursun}.} \bibinfo{year}{[n. d.]}\natexlab{}.
\newblock \bibinfo{booktitle}{{\em {Advanced Data Mining Techniques}}}.
\newblock \bibinfo{publisher}{Springer}.
\newblock
\showISBNx{9783540769163}
\showDOI{%
\url{https://doi.org/10.1007/978-3-540-76917-0}}


\bibitem[\protect\citeauthoryear{{OPC Foundation}}{{OPC Foundation}}{2017}]%
        {OPCFoundation.2017}
\bibfield{author}{\bibinfo{person}{{OPC Foundation}}.}
  \bibinfo{year}{2017}\natexlab{}.
\newblock \bibinfo{title}{Unified Architecture}.
\newblock   (\bibinfo{year}{2017}).
\newblock
\showURL{%
\url{https://opcfoundation.org/developer-tools/specifications-unified-architecture/part-1-overview-and-concepts}}


\bibitem[\protect\citeauthoryear{PROFIBUS}{PROFIBUS}{2017}]%
        {PROFIBUS.2017}
\bibfield{author}{\bibinfo{person}{PROFIBUS}.} \bibinfo{year}{2017}\natexlab{}.
\newblock \bibinfo{title}{PROFINET Specification}.
\newblock   (\bibinfo{year}{2017}).
\newblock
\showURL{%
\url{http://www.profibus.com/nc/download/specifications-standards/downloads/profinet-io-specification/display/}}


\bibitem[\protect\citeauthoryear{Project}{Project}{[n. d.]}]%
        {Bro.}
\bibfield{author}{\bibinfo{person}{The~Bro Project}.} \bibinfo{year}{[n.
  d.]}\natexlab{}.
\newblock \bibinfo{title}{{The Bro Network Security Monitor}}.
\newblock   (\bibinfo{year}{[n. d.]}).
\newblock
\showURL{%
\url{https://www.bro.org/}}


\bibitem[\protect\citeauthoryear{{Rapid7}}{{Rapid7}}{[n. d.]}]%
        {metasploit.}
\bibfield{author}{\bibinfo{person}{{Rapid7}}.} \bibinfo{year}{[n.
  d.]}\natexlab{}.
\newblock \bibinfo{title}{{metasploit}}.
\newblock   (\bibinfo{year}{[n. d.]}).
\newblock
\showURL{%
\url{https://www.metasploit.com/}}


\bibitem[\protect\citeauthoryear{Rokach and Maimon}{Rokach and Maimon}{2005}]%
        {Rokach.2005}
\bibfield{author}{\bibinfo{person}{Lior Rokach} {and} \bibinfo{person}{Oded
  Maimon}.} \bibinfo{year}{2005}\natexlab{}.
\newblock \showarticletitle{Top-Down Induction of Decision Trees
  Classifiers—A Survey}.
\newblock \bibinfo{journal}{{\em IEEE Transactions on Systems, Man, and
  Cybernetics, Part C (Applications and Reviews)\/}} \bibinfo{volume}{35},
  \bibinfo{number}{4} (\bibinfo{date}{November} \bibinfo{year}{2005}),
  \bibinfo{pages}{476--487}.
\newblock
\showISSN{1094-6977}
\showDOI{%
\url{https://doi.org/10.1109/TSMCC.2004.843247}}


\bibitem[\protect\citeauthoryear{Rousseeuw}{Rousseeuw}{1987}]%
        {Rousseeuw.1987}
\bibfield{author}{\bibinfo{person}{Peter~J. Rousseeuw}.}
  \bibinfo{year}{1987}\natexlab{}.
\newblock \showarticletitle{{Silhouettes: A graphical aid to the interpretation
  and validation of cluster analysis}}.
\newblock \bibinfo{journal}{{\it J. Comput. Appl. Math.}}  \bibinfo{volume}{20}
  (\bibinfo{date}{November} \bibinfo{year}{1987}), \bibinfo{pages}{53--65}.
\newblock
\showDOI{%
\url{https://doi.org/10.1016/0377-0427(87)90125-7}}


\bibitem[\protect\citeauthoryear{{Schneider Electric}}{{Schneider
  Electric}}{2017}]%
        {Schneider-Electric.2017}
\bibfield{author}{\bibinfo{person}{{Schneider Electric}}.}
  \bibinfo{year}{2017}\natexlab{}.
\newblock \bibinfo{title}{{Life Is On}}.
\newblock   (\bibinfo{year}{2017}).
\newblock
\showURL{%
\url{https://www.schneider-electric.fr/fr/}}


\bibitem[\protect\citeauthoryear{Seidl}{Seidl}{[n. d.]}]%
        {Seidl.2015}
\bibfield{author}{\bibinfo{person}{Jan Seidl}.} \bibinfo{year}{[n.
  d.]}\natexlab{}.
\newblock \bibinfo{title}{{VirtuaPlant}}.
\newblock   (\bibinfo{year}{[n. d.]}).
\newblock
\showURL{%
\url{https://wroot.org/posts/introducing-virtuaplant-0-1/}}


\bibitem[\protect\citeauthoryear{Siaterlis, Genge, and Hohenadel}{Siaterlis
  et~al\mbox{.}}{2013}]%
        {Siaterlis.2013}
\bibfield{author}{\bibinfo{person}{Christos Siaterlis}, \bibinfo{person}{Bela
  Genge}, {and} \bibinfo{person}{Marc Hohenadel}.}
  \bibinfo{year}{2013}\natexlab{}.
\newblock \showarticletitle{{EPIC: A Testbed for Scientifically Rigorous
  Cyber-Physical Security Experimentation}}.
\newblock \bibinfo{journal}{{\em IEEE Transactions on Emerging Topics in
  Computing\/}} \bibinfo{volume}{1}, \bibinfo{number}{2}
  (\bibinfo{date}{December} \bibinfo{year}{2013}), \bibinfo{pages}{319--330}.
\newblock
\showISSN{2168-6750}
\showDOI{%
\url{https://doi.org/10.1109/TETC.2013.2287188}}


\bibitem[\protect\citeauthoryear{Snort}{Snort}{[n. d.]}]%
        {Snort.}
\bibfield{author}{\bibinfo{person}{Snort}.} \bibinfo{year}{[n. d.]}\natexlab{}.
\newblock \bibinfo{title}{{Snort}}.
\newblock   (\bibinfo{year}{[n. d.]}).
\newblock
\showURL{%
\url{https://www.snort.org/}}


\bibitem[\protect\citeauthoryear{{The R Foundation}}{{The R Foundation}}{[n.
  d.]}]%
        {R.}
\bibfield{author}{\bibinfo{person}{{The R Foundation}}.} \bibinfo{year}{[n.
  d.]}\natexlab{}.
\newblock \bibinfo{title}{{The R Project for Statistical Computing}}.
\newblock   (\bibinfo{year}{[n. d.]}).
\newblock
\showURL{%
\url{https://www.r-project.org/}}


\bibitem[\protect\citeauthoryear{{The University of Utah}}{{The University of
  Utah}}{[n. d.]}]%
        {Emulab.}
\bibfield{author}{\bibinfo{person}{{The University of Utah}}.}
  \bibinfo{year}{[n. d.]}\natexlab{}.
\newblock \bibinfo{title}{{emulab total network testbed}}.
\newblock   (\bibinfo{year}{[n. d.]}).
\newblock
\showURL{%
\url{http://www.emulab.net/}}


\bibitem[\protect\citeauthoryear{TU Wien}{TU Wien}{2017}]%
        {e1071.2017}
TU Wien \bibinfo{year}{2017}\natexlab{}.
\newblock \bibinfo{booktitle}{{\em {Package e1071}}}.
\newblock TU Wien, Probability Theory Group (Formerly: E1071).
\newblock


\bibitem[\protect\citeauthoryear{University of Berkeley}{University of
  Berkeley}{2015}]%
        {randomForest.2015}
University of Berkeley \bibinfo{year}{2015}\natexlab{}.
\newblock \bibinfo{booktitle}{{\em {Package randomForest}}}.
\newblock University of Berkeley.
\newblock


\bibitem[\protect\citeauthoryear{{University of California, Irvine
  (UCI)}}{{University of California, Irvine (UCI)}}{1999}]%
        {KDD.1999}
\bibfield{author}{\bibinfo{person}{{University of California, Irvine (UCI)}}.}
  \bibinfo{year}{1999}\natexlab{}.
\newblock \bibinfo{title}{{KDD Cup 1999 Data}}.
\newblock   (\bibinfo{year}{1999}).
\newblock
\showURL{%
\url{http://kdd.ics.uci.edu/databases/kddcup99/kddcup99.html}}


\bibitem[\protect\citeauthoryear{van Rijsbergen}{van Rijsbergen}{1979}]%
        {Rijsbergen.1979}
\bibfield{author}{\bibinfo{person}{C.~J. van Rijsbergen}.}
  \bibinfo{year}{1979}\natexlab{}.
\newblock \bibinfo{title}{Information Retrieval}.
\newblock   (\bibinfo{year}{1979}).
\newblock


\bibitem[\protect\citeauthoryear{Wang, Fang, and Dai}{Wang
  et~al\mbox{.}}{2010}]%
        {Wang.2010}
\bibfield{author}{\bibinfo{person}{Chunlei Wang}, \bibinfo{person}{Lan Fang},
  {and} \bibinfo{person}{Yiqi Dai}.} \bibinfo{year}{2010}\natexlab{}.
\newblock \showarticletitle{{A Simulation Environment for SCADA Security
  Analysis and Assessment}}. In \bibinfo{booktitle}{{\em {2010 International
  Conference on Measuring Technology and Mechatronics Automation}}},
  Vol.~\bibinfo{volume}{1}. \bibinfo{pages}{342--347}.
\newblock
\showISSN{2157-1473}
\showDOI{%
\url{https://doi.org/10.1109/ICMTMA.2010.603}}


\bibitem[\protect\citeauthoryear{Yang, Usynin, and Hines}{Yang
  et~al\mbox{.}}{2006}]%
        {Yang.2006}
\bibfield{author}{\bibinfo{person}{Dayu Yang}, \bibinfo{person}{Alexander
  Usynin}, {and} \bibinfo{person}{J.~Wesley Hines}.}
  \bibinfo{year}{2006}\natexlab{}.
\newblock \showarticletitle{{Anomaly-based intrusion detection for SCADA
  systems}}. In \bibinfo{booktitle}{{\em 5. International Topical Meeting on
  Nuclear Plant Instrumentation Controls, and Human Machine Interface
  Technology}}. \bibinfo{pages}{797--803}.
\newblock


\bibitem[\protect\citeauthoryear{Zhu, Joseph, and Sastry}{Zhu
  et~al\mbox{.}}{2011}]%
        {Zhu.2011}
\bibfield{author}{\bibinfo{person}{Bonnie Zhu}, \bibinfo{person}{Anthony
  Joseph}, {and} \bibinfo{person}{Shankar Sastry}.}
  \bibinfo{year}{2011}\natexlab{}.
\newblock \showarticletitle{{A Taxonomy of Cyber Attacks on SCADA Systems}}. In
  \bibinfo{booktitle}{{\em 2011 International Conference on Internet of Things
  and 4th International Conference on Cyber, Physical and Social Computing}}.
  \bibinfo{pages}{380--388}.
\newblock
\showDOI{%
\url{https://doi.org/10.1109/iThings/CPSCom.2011.34}}


\end{thebibliography}

\end{document}